# *SKALA, a log-periodic array antenna for the SKA-low instrument: design, simulations, tests and system considerations

E. de Lera Acedo[1], N. Razavi-Ghods[1], N. Troop[2], N. Drought[2], and A.J. Faulkner[1]

[1]Cavendish Laboratory, Department of Physics, University of Cambridge, Cambridge CB3 0HE, UK
email: eloy@mrao.cam.ac.uk

[2]Cambridge Consultants Ltd., Cambridge, UK

The very demanding requirements of the SKA-low instrument call for a challenging antenna design capable of delivering excellence performance in radiation patterns, impedance matching, polarization purity, cost, longevity, etc. This paper is devoted to the development (design and test of first prototypes) of an active ultra-wideband antenna element for the low-frequency instrument of the SKA radio telescope. The antenna element and differential low noise amplifier described here were originally designed to cover the former SKA-low band (70-450MHz) but it is now aimed to cover the re-defined SKA-low band (50-350MHz) and furthermore the antenna is capable of performing up to 650MHz with the current design. The design is focused on maximum sensitivity in a wide field of view (+/- 45° from zenith) and low cross-polarization ratios. Furthermore, the size and cost of the element has to be kept to a minimum as millions of these antennas will need to be deployed for the full SKA in very compact configurations. The primary focus of this paper is therefore to discuss various design implications for the SKA-low telescope.

## 1. INTRODUCTION

The Square Kilometre Array, SKA [1], will be the next generation metre and centimetre radio telescope and is currently under development. The full SKA will have 100x more sensitivity and able to search the sky a million times faster than any current radio telescope. It is a major international project, which is expected to cost more than €2billion. The SKA will be built in remote areas in Australia and Southern Africa where there is minimum radio interference for the best performance. The SKA has major science goals for its lowest frequency collector system covering 50 to 350MHz [2]. One of the main science experiments is to observe neutral hydrogen in the very early universe, the Epoch of Reionisation. Here the emission from hydrogen at 1420MHz in the rest frame has been red shifted to sub 100MHz frequencies. The required sensitivity to detect such a weak signal at these frequencies requires a very large collecting area in order to overcome the high sky brightness temperatures below ~150MHz. The most practical way to achieve this is by using large number of phased array antennas.

The development of the SKA is technically very challenging for most aspects of the system: processing, data volumes, cost etc. The low frequency phased array is no exception. The element and array design needs very careful and innovative thinking in order to meet the specification. The SKA requirements include: high sensitivity, excellent polarization performance, more than ±45° scan angles, low receiver noise above 150MHz, stability and, of course, low cost and power [3]. Furthermore, the maximum dimensions of the antenna needed to be constrained such that elements could be mounted on a ~1.5m pitch [3]. To reduce the overall system cost including antennas, signal processing, signal transport and central processing a single element covering the band is highly beneficial. This makes a very challenging set of design requirements.

The SKA will be built in two phases; Phase 1 is scheduled for procurement and construction 2017 – 2023, with the full SKA in the 2020s. To meet the sensitivity requirements approximately 250,000 elements will be required for Phase 1 rising to >3 million in the full SKA. It is therefore essential to have an optimized and fully tested design prior to commencing construction. The first challenge is the design of a dual-pol element that is capable of performing electromagnetically over the full frequency range maintaining wide beamwidth with high polarization isolation. The log periodic antenna described here meets the requirements and provides the basis for detailed development for deployment in the SKA. Generally, log periodic antennas would not be considered to have a wide enough beamwidth, however, one can optimize the design aiming for high sensitivity at large scan angles.

The design of a differential low noise amplifier (LNA) for minimum receiver noise is also discussed here. This involves providing an optimal noise match to the antenna impedance. In the present design, the LNA is located at the feeding point of the antenna for optimum noise performance. The LNA, for each polarization, consists on a pair of identical low noise transistors directly feeding each arm of the antenna and are closely followed by a wideband balun to provide a single ended output.

With a large number of elements, low manufacturing and deployment costs are essential. Here the initial approach at "design for manufacture" is considered, the fundamental structure lends itself to this task. Ease of alignment is critical, with the required precision to achieve SKA requirements being part of the simulation process. The SKA is specified to be operational for at least 30 years, while upgrades will take place, the front end collector technology will be designed to survive in desert conditions for at least this length of time.

In Section 2 we discuss the design targets and strategy for the antenna element as well as the simulated analysis. The implications of a specific antenna design to the overall system design are also included in this section. Section 3 presents the first mechanical design for low cost production. Section 4 is then dedicated to describing the low noise amplifier design. Section 5 discusses the different tests that we have done to both the antenna and LNA as single elements and in the array environment in 16-element prototype arrays. Finally we conclude with a Section 6 describing the conclusions of the paper.

## 2. Antenna design and Simulations

### A. Design

The SKA telescope will be dived in 2 main instruments. An aperture array type instrument covering frequencies from 50 MHz up to at least 1.4 GHz and a dish-based interferometer covering a frequency band up to at least 10 GHz. The lower part of the band of the aperture array instrument will be called SKA-low (this instrument will be in Western Australia). The Low-Frequency Aperture Array (LFAA) is defined as the set of antennas, on board amplifiers and local processing required for the low frequency Aperture Array telescope of the SKA. The design of the antenna targeted the specifications required to deliver the electromagnetic performance required for the SKA-low telescope [3]. The current antenna was originally designed to work from at least 70 to 450 MHz. The SKA baseline design requires an element operating in the band 50-350MHz. The main science goal in this frequency band is the study of the Epoch of Re-ionization (EoR) [2]. One of the main figures of merit for a radio telescope is the Survey Speed. This is directly proportional to the Field of View (*FoV*), the channel bandwidth (dependent on the signal processing system) and the sensitivity squared. In [4-7] studies of the array configuration have shown how sparse-irregular configurations are a good choice for low frequency radio telescopes, especially if we would like to to maximize sensitivity. These advantages include the reduction of grating lobes into a higher average side lobe level [4-6] and the randomization of mutual coupling effects [7]. In [8] one can also read about an optimization strategy specifically designed for these types of arrays. This is the choice for SKA-low [3]. A brief description of the main design parameters for the design of an array antenna for low frequency radio astronomy follows.

1) **The sensitivity** in radio astronomy is proportional to the telescope's **effective aperture area over the system noise temperature** (*A/T*) (see equation 1). Both of these terms are dependent on the antenna front end [9,10]. Therefore, the following parameters need to be optimized for a phased array antenna:
   - **a. Radiation efficiency ($\eta_{rad}$).**
     It does affect both the incoming desired wave, by weighting the Directivity ($D_{\theta,\Phi}$) transforming it into Gain ($G_{\theta,\Phi}$),, and the system noise. The definition of Gain in here does not include the effect of impedance mismatch (*IEEE* definition). It affects the system noise in several ways: by directly weighting the undesired Sky noise temperature as well as representing the losses at the antenna terminals. These losses are due to the presence of non-perfect electric conductors in the antenna and the presence of ground soil, both at the physical temperature ($T_0$).
   - **b. Antenna temperature, $T_A$.**
     The antenna temperature, which is a function of the array antenna gain, depends on several factors related to the sky noise temperature and is a limiting factor at the low end of the band.
   - **c. Receiver temperature ($T_{rec}$)**
     This factor can dominate the system temperature and therefore the sensitivity at the high end of the band, where the sky noise is no longer dominant [10].

**d. Antenna gain within the field of view**

The antenna gain describes the spatial discrimination capabilities of the antenna, which has a direct impact on affective area and thus is one of the main factors in limiting the *A/T* of the system. As mentioned above, in this paper we follow the *IEEE* definition for the Gain (*The ratio of the radiation intensity, in a given direction, to the radiation intensity that would be obtained if the power accepted by the antenna were radiated isotropically*).

$$\left.\frac{A_{\mathrm{eff}}}{T_{sys}}\right|_{\theta.\varphi} = \frac{\lambda^2/_{4\pi} \cdot G_{\theta.\varphi}}{\eta_{\mathrm{rad}} \cdot T_{\mathrm{A}} + \left(1-\eta_{\mathrm{rad}}\right)\cdot T_0 + T_{\mathrm{rec}}} \tag{1}$$

2) Low relative **cross-polarization** level is required in order to be able to measure the foregrounds of the EoR [2]. Also, it is important for the telescope calibration. The required specification levels for the SKA-low science cases are currently under analysis. In the context of wide field radio astronomy, cross-polarization has been in the last few years assessed using the Intrinsic Cross Polarization ratio (IXR) [11] (see equation 3), which is a measure of the condition number of the Jones matrix (*J*) relating the projection of the field components in the sky (*E*) to the voltage signals at the output of the antenna terminals for both polarizations (*V*). It should be pointed out, however, that an IXR characterization requires far-field vector measurements. This of course imposes a limitation for low frequency arrays of the size of SKA stations as this can only be accurately done currently for single antennas or small arrays. This is good enough for the design process but imposes a limitation for large arrays where astronomical measurements will need to be further developed to provide the desired accuracy.

$$\begin{bmatrix} V_1 \\ V_2 \end{bmatrix} = \begin{bmatrix} J_{11} & J_{12} \\ J_{21} & J_{22} \end{bmatrix} \begin{bmatrix} E_t \\ E_p \end{bmatrix} \tag{2}$$

$$IXR_J = \left(\frac{\kappa(J)+1}{\kappa(J)-1}\right)^2 \tag{3}$$

3) **The element's footprint and the array configuration**. Nearly 75% of the SKA stations will reside in the core area where they will form a sea of antenna elements. The division of this core in stations will be done therefore in the data processing and these will be logical rather than physically divided stations. First of all we need to point out that the antenna grid (station configuration) cannot be regular due to detrimental mutual coupling effects. These include impedance anomalies across the frequency range as well as strong dips in the pattern of the embedded elements. In Fig. 1 we can see a plot of the $S_{11}$ for the antenna element embedded in an infinite regular array when all elements are active and pointing at zenith, showing clearly the type of anomalies to be expected for such a configuration. Also, as described in [7], only if the configuration is randomized the effects of mutual coupling are as well randomized allowing us to achieve the desired sensitivity. In [12] the effect of the station size in this randomization is described, from where we can conclude that 256 elements are still a good number. As described in [4-6], the effects of the side lobes in irregular/randomize configurations tend to be less detrimental. Figure 2 shows an example of a random configuration for a 35 m station with 256 antenna elements as we may have for SKA. Furthermore, in order to deliver maximum brightness sensitivity; the filling factor of the SKA core must be as high as possible [2]. In the core all the elements contribute to capture the relevant Fourier modes in a given angular scale. This can be pictured as all the antennas being cross-correlated to each other in an interferometry sense, so the distance between the antennas inside a station is as well a baseline contributing to the total sensitivity. Therefore a higher filling factor will maximize the information collected from the sky. This means that the foot-print needs to be kept to a minimum. Furthermore, the SKA calls for a station beam-width of 5 degrees at ~100 MHz (middle of the EoR frequency band) as described in [2] delivering a station diameter of 35 m. It is also stated in the latest requirements of the telescope that each of these stations will contain 256 elements. Therefore, according to the latest SKA specifications the average area allocated per element should then equal 3.75 $m^2$. This are gives an approximate square foot print per element of 1.93x1.93m, which consequently defines the transition between the dense and sparse regimes [10]. Therefore, in order for the antenna elements to be positioned in such a random grid, the footprint of the antenna needs to be smaller than 1.93x1.93m. The current design targeted a minimum separation of 1.5 m, which would allow achieving a configuration where the array side lobes have been redistributed into a sea of side lobes [5] and effects of mutual coupling have averaged out [7]. Mutual coupling is a crucial design consideration for any element working in an array environment as it modifies the element pattern and scan impedance [10]. In our particular case, backed up by our studies of mutual coupling of irregular sparse arrays suitable for SKA-low [7,10,12], we have implemented the first design process of the array antenna using a single element. The idea behind this is that in a randomized array the mutual coupling effects tend to randomize out and the array pattern can be approximated to a first order by the multiplication of the array factor and the single element pattern (equal for all antennas elements). It is worth noticing however that in the calibration step mutual coupling effects will need to be included. Some related work to this can be found in [12,13].

4) **Antenna field of view**. The SKA calls to cover a region of the sky +/- 45$^o$ from zenith [3]. The current SKA requirements don't establish a value such as 3 dB half power beam width across the frequency range but rather expects the sensitivity at all angles within that range to be maximized. The *FoV* of each element must therefore be +/- 45$^o$ from zenith and will determine the maximum *FoV* of the instrument, since the elements will be electronically beam-formed/correlated. A very directive antenna will show great sensitivity at zenith but will underperform at large angles from zenith. Furthermore, a low directivity antenna will have similar values of *A/T* at all angles within the *FoV* but will not be able to deliver the required *A/T* at zenith (see [3]). One of the key concepts behind the LFAA antenna system is the use of antennas that maximize sensitivity within the *FoV*. It is therefore worth noticing that it is not the half power beamwidth that matters here, but the *A/T* delivered at a given scan angle. Therefore, a slightly more directive element than one with a half-power beamwidth of 90$^o$ may, in turn, deliver a better overall sensitivity across most of the FoV at the expense of a small reduction near the edges of the *FoV* (see Fig. 3). The SKALA element has been optimized to maximize this trade-off.

5) **The cost per element.** The full SKA will have more than 3 million of these antennas, it is therefore essential to reduce the cost of production and deployment per element to the minimum.

6) **Durability.** The hard conditions in the Western Australian desert require a careful choice of materials capable of surviving for more than 30 years.

With these considerations, a Log-Periodic Dipole Array (LPDA), named *SKALA* (*SKA Log-periodic Antenna*) has been designed and optimized to meet all the listed requirements. The basic design of this element is presented in [14]. It consists on a 9 dipole LPDA designed to be fed by a differential LNA, for minimum noise in order to achieve maximum sensitivity in the desired FoV (see Fig. 4). The top dipoles of SKALA are not resonant in the original design band but they serve as dummy dipoles to preserve the ultra-wideband performance of the element. Furthermore, the bottom dipole was modified as shown in Fig. 5 to improve the impedance matching to the LNA at low frequencies while reducing the footprint of the antenna. The top image on the left of Fig. 5 shows the initial design with a straight bottom dipole and the bottom image shows the modification. This modification has transformed the bottom dipole into a type of bow-tie radiating structure, which improves the impedance matching while reducing the footprint of the antenna. Figure 6 shows a view of the current version of SKALA including the mechanical improvements described in Section 3 and Fig. 7 shows the final feeding for simulations including the LNA boards, a differential port for the excited polarization (reference impedance = 100 Ω) and 2 lumped elements for the non-excited polarization. Each of these lumped elements represents the input impedance of the transistors forming the pseudo-differential LNA (see Fig. 7). Table 1 shows the main dimensions of SKALA's computer model for the final design. It has also been decided that in order to maintain a similar environment for all the antennas in the SKA for calibratability reasons and in order to be more isolated from the soil conditions, a metallic ground screen will be used under the antenna elements. The antenna has a footprint of 1.2x1.2m and it is 1.8m in height.

B. Simulations

The simulated results presented in this section are the produce of several parametric analyses on the LPDA design. Unless otherwise stated all the simulations are performed with an infinite metallic ground plane under the antenna. It is worth pointing out that a numerical method for simulation of large irregular arrays such as the ones needed for the SKA is being developed and it is based on the technique described in [15]. This Method of Moments (MoM) technique makes use of Macro Basis Functions (MBFs), which correspond to an a-priori determined set of possible current distributions on the antennas. The interactions between MBFs on different antennas are obtained in an extremely efficient way thanks to a physically-based preliminary mapping versus relative positions between the antennas. The errors resulting from the use of MBFs and their fast mapping are between -30 dB and -40 dB below the solution. This technique can be applied to any particular configuration of antennas making it an ideal choice to solve EM problems such the one presented by SKA stations, where we have arrays of disconnected antennas. The simulations presented in this section, unless otherwise stated have been carried out using the commercial software CST. Furthermore, simulations of lightning effects are also being carried out in order to design appropriate protection systems [16].

1) **Impedance**. Figure 8 shows the impedance of one of the log-periodic pair of arms of SKALA when the orthogonal pair of arms is loaded with 100 **Ω.** The parametric analysis explored variations on the main parameters affecting the impedance of the antenna: width of the transmission line, *W*; length of the transmission line, *L*; logarithmic growth ratio, $\tau$; width of the dipoles; width of the feed, *G*; opening angle, $\alpha$; and the distance to ground, *gnd_D*. In Fig. 6 we can see the impedance of both the basic design of Fig. 4 and of the final design (including the modified bottom arm, the wire dipoles and the feed optimization for matching with the LNA). This optimization lead to an improvement of the receiver noise temperature of up to 23% at spot frequencies across the band. It is important to note here (as it was described in [10]) that due to the high sky noise temperatures at the low end of the band, the antenna impedance requirements can be relaxed, as the receiver temperature will only be a small fractional part of the total system noise temperature in a radio telescope at these frequencies. This is the reason why the visible resonance at the low end of the band does not show a strong effect on the total system noise temperature (see Fig. 9).

2) **Pattern**. Figures 10-12 show the isolated antenna pattern at 4 frequencies across the band of interest (70, 150, 300 and 450MHz) for H- and E-plane as well as 45$^{o}$ cuts. The patterns are shown for the single polarized antenna over an infinite ground plane, over a typical soil and over a metallic mesh over soil. The orthogonal arms are present and loaded with 100 Ω. The patterns are simulated in a transmit mode where the active polarization is excited with a discrete port in CST (signal source in series with a 100 Ω load). The soil is a representation of Australian soil [17] and has 10 % moisture content, a relative permittivity of 6.31 and a conductivity of 22 mS/m. The metallic wire mesh has a pitch of 30 cm and a wire thickness of 2.5 mm. It can be appreciated that the effect on the pattern of the soil and the mesh, due to the directive nature of the LPD antennas, is only significant at low frequencies while it is almost unnoticed at high frequencies. It can also be seen that the back lobe of the antenna quickly decays with increasing frequency, reducing the noise picked up from the ground by SKALA.

3) **Cross-polarization** (IXR). The IXR, computed as in equation 3, is presented in Figure 13 for the case when the antenna is on top of an infinite metallic ground plane. Detailed simulations are being carried out in order to establish the minimum IXR needed for the SKA-low [18,19]. Furthermore, several studies have analyzed the misalignment tolerances of array antennas for low frequency radio astronomy applications [20,21]. The simulation results presented here show that IXR levels higher than 15 dBs are possible across the whole *FoV* and frequency range. Furthermore, simulations of IXR where the antenna is over an infinite metallic ground plane compared to when it is or over a typical soil showed differences of ~ 10 dB at zenith below 100 MHz (in favor of the infinite ground plane case). Also, using a metallic ground plane, the variations between different antennas in the array will be reduced (as the antennas won't be so sensitive to soil conditions) facilitating their modeling and calibration.

4) **Radiation efficiency**. Figure 14 shows the radiation efficiency both with soil and with a ground mesh in between the antenna and the soil. The radiation efficiency is especially important at the high end of the frequency band, where the sky noise temperature does not dominate the noise performance of the system [10]. The results show that the radiation efficiency nearly reaches 100% (~97%) at about 200 MHz. The radiation efficiency here is calculated including the presence of the soil and not only the antenna structure. The losses due to the absorption of the soil are therefore included in the radiation efficiency. This is the reason why the radiation efficiency is low at the lower end of the frequency range, where the antenna is more sensitive to the presence of the soil (the current distribution is mostly confined in the bottom dipoles). The inclusion of the effect of the soil in the calculation of the radiation efficiency is justified by the calculation of sensitivity as described by equation (1).

5) **Sensitivity** (*A/T*). The sensitivity is calculated as described in equation 1. In the plot shown in Fig. 15, the per-polarization *A/T* for the $SKA_1$ telescope based on SKALA elements. The calculations in this paper are based on a $T_A$ $(T_{sky})$ as in equation (4), which is the same uniform sky assumed in the baseline design document [3]. This is of course an average value, which does not correspond with a specific observation. A specific observation of, for example, a cold patch of the sky would deliver different *A/T* values, due to the changing station beam on the sky while the array is scanned. But by using the sky noise defined by a uniform model such as the one in equation (4) we can have an idea of the average *A/T* of the instrument.

$$T_{sky} = 60 \cdot \lambda^{2.55} \quad (4)$$

The calculation leading to the simulated values of sensitivity presented here assumes $2^{18}$ SKALA antennas in 1024 arrays with average spacing of 1.93 m between antenna elements (256 elements in 35 m stations). This means that because of the effect of the mutual coupling and the radiation efficiency, the effective aperture is reduced at 50 MHz and ultimately limited by the physical size of the antenna unit cell (1.93 x 1.93 m) at the low end of the frequency band [22]. The antennas are backed by a metallic mesh grid with a 30 cm pitch. The soil below that mesh plane has moisture content of 10% and it is typical of the Australian desert and this has been included in the electromagnetic simulations. The receiver noise is based on the current design of the LNA. Figure 14 shows the sensitivity at zenith and at 30° and 45° scan. The off-zenith sensitivity values are presented here as a first order average of the E- and H-plane sensitivities, assuming an identical orientation for all the antennas in the array.

## 3. Mechanical design

More than 3 million antenna elements will be needed for the final SKA and they must survive for at least 30 years in the Australian desert, so an inexpensive and durable design is critical. In collaboration with a UK company, Cambridge Consultants Ltd we have developed an antenna construction based on formed steel wires which are then welded to a tubular steel spine. The whole structure is then supported by a glass-fiber reinforced plastic (GFRP) tube. The manufacturing cost of the antenna parts (including LNA electronics and ground plane mesh) is targeted to be approximately 75 Euro, at production rates exceeding 100,000 per annum. This includes either coating of the metal parts or the use of stainless steel for long durability.

As shown in Figure 5 the first prototype antenna design consisted of four flat arms cut from 2mm thick steel sheets. This uses a large area of material and weighed 5.2 kg per panel. Since the central portion of each dipole is substantially passive from an electromagnetic perspective it may be removed to save material cost, weight and wind loading without loss of performance. It is a well-known target of design for manufacture and assembly (DFMA) that cost reduction is driven by reducing parts count, material usage & assembly operations. Several options were studied [23]. The construction option selected uses nine U-shaped loops welded to a tubular spine. This allows thinner wire to be used for the shorter arms, and the least number of bending and welding operations. The U-shaped loops can each be made using simple bending equipment. With this configuration, all eighteen joints can be welded in the one setup, manually or robotically. The steel wire and tube fabrication of the arm shown in Fig. 16 (current SKALA antenna design) weighs approximately 1.6 kg each.

The fabricated steel sub-assemblies can be flat-packed for efficient and easy transport to the installation site, along with the GFRP composite tube and additional support parts. The latter consists of a 2-part or 3-part moulded central support structure that can be clamped to the GFRP tube, and a two-part moulded enclosure to protect the electronics – the lower half of which may be permanently joined to the GFRP tube to form another factory built sub-assembly. The size and internal details of the moulded enclosure may be varied to accommodate different sets of electronics, e.g. using conventional RF coax cable output with phantom DC power input, or using a local power source (such as solar cells) and analogue over fibre RF output. Both approaches for power & signal connection may use the same mechanical design.

In the current design presented in this paper, the GFRP tube may be attached to a variety of anchors or foundations that suit local ground or soil conditions, and surrounded by the ground plane mesh. The antenna is still under development and the next version may likely include a base for support so that the central tube does not need to be clamped to the ground. The central support structure is then attached to the tube and each of the four fabricated panels can be fitted using 3 points of attachment. The RF cables need to be routed inside the GFRP tube within the antenna structure but they can exit the tube through a slot just above ground level. The cables may either be factory fitted or fitted on site depending on the length required, so that the number of coaxial RF connectors can be minimised in order to save cost and RF losses.

In order to achieve a resonance free band response due to for example common-modes, symmetry is a very desired feature at the feeding point of the antenna. At the top of the antenna (see Fig. 17), there are 3 planes that might allow symmetrical connections (horizontal, vertical-transverse and vertical-diagonal). The most convenient arrangement for SKALA is based on using two identical single-sided PCBs mounted horizontally, but with the coax output cable connector fitted from opposite sides (see Fig. 17).

## 4. Low Noise Amplifier design

The LNA was envisaged as a pseudo-differential amplifier (see Fig. 18) in order to achieve the minimum possible noise temperature [24], which is a critical parameter in order to obtain maximum sensitivity. The initial target for the noise figure was set to 0.4 dBs at the high end of the band and the target gain was set to be higher than 40 dBs in order to reduce the contribution to the system noise from any other component of the receiving chain. At this stage the LNAs was designed to be powered over the signal coax (the so-called Phantom Power method). If the LNA is to operate with an optimum Noise Figure over a range of frequencies, a wideband impedance transformation is required. In this application the impedance transformation is addressed by considering the design of the antenna and transmission line in such a way that the optimal impedance presented to the LNA is a close to optimal as is possible with greater weight given to achieving $\Gamma_{opt}$ at the high frequency end of the band where the sky noise temperature is lowest. Furthermore, the presence of these boards at the feeding point of the antenna should not alter the antenna performance and needs to be taken into account in the antenna design and optimization. Not only will this affect the noise performance but also the rejection of common modes and the LNA gain amongst others. It is therefore essential for a radio astronomy application where high specs are needed to realize the design of both the antenna and LNA in conjunction. In order to preserve a high level of common mode rejection a dual active device was chosen (Avago MGA-16516 [25]). The differential signal was then applied to a well-balanced wideband transformer which converted the signal in single ended form whilst preserving CMRR. The single ended signal was subsequently amplified with a second low noise gain stage in order to obtain the required overall gain from the amplifier module. Normally an LNA exposed to such a spectrally rich range of input signals would have to have a very high IP3 in order that intermodulation products do not limit its performance. However this is not such as problem as it might at first seem as the low frequency part of the SKA is to be deployed in the desert of Western Australia which has a suitably low level of radio occupancy as to make low frequency wideband radio astronomy feasible.

A simulation of LNA noise figure with its input terminated in the antenna impedance is shown in Figs. 19 and 20. Actual noise figure is degraded above the $F_{min}$ value by the fact that each device in the MGA-16516 is a non-correlated noise source and the there is a noise contribution from the second stage LNA, which is small due to the high gain of the first stage. The simulated transducer gain of the LNA with its input port terminated with the antenna impedance is shown in Fig. 21 and is relatively flat across band. In practice, variation in system gain will be taken into account by calibration. Figure 22 shows a view of the LNA without the screening can and the plastic spacer that separates the top board (polarization X) from the bottom board (polarization Y).

## 5. MEASUREMENTS

### A. Single antenna measurements

Multiple measurements have been performed on the SKALA antenna. These tests have been performed on scaled prototypes of SKALA [23] as well as on the real size prototypes. Figure 23, shows a measurement of the antenna impedance using a differential measurement technique [24,26] and its comparison with simulations. The measurement of differential devices is always challenging as most test equipment is designed for single-ended devices. It is therefore necessary to design specific jigs and feeding boards to perform the required tests which then need to be de-embeded from the measurements. These results show a very good agreement confirming the validity of the simulation work. Figure 24 shows the pattern measurements again validating the simulations. These measurements were done with the antenna on top a 2x2 m ground plane. Due to the large wavelengths at these frequencies these measurements were performed in an outdoor ground reflection test range at QinetiQ, UK. In these measurements, after finding the optimal vertical position of both the test antenna and the SKALA element, a test log-periodic antenna was used to transmit a CW signal while one of the polarizations of the SKALA element was used to receive it. The orthogonal arms were loaded with 100 Ω. The SKALA element was then rotated (the center of rotation was the center of the ground plane) in order to obtain the pattern cut. These tests also confirmed the good agreement between measurements and simulations. The measurement of radiation patterns at these low frequencies is not trivial and measurements at anechoic chambers is typically limited to frequencies above 300MHz. Currently we are exploring other possibilities such as the use of flying vehicles [27] to perform these tests for the new versions of SKALA. The complexity on testing our prototypes meant that we had to spend a large portion of the development process designing tests for the validation of the different simulation software packages used for the array antenna design. Only a very limited set of measurements can really be practically obtained for these large systems at these low frequencies but it is crucial to confirm that the simulation tools are working so that the design work can continue using the simulations as main input.

### B. Amplifier measurements

A number of tests were carried out to evaluate the LNA module gain, noise figure and intermodulation performance. For gain measurements, a VNA was used along with a hybrid splitter to feed the differential inputs to the board and the loss of the splitter was then de-embedded from the results (see Fig. 25). Measurements of the transducer gain (when connected to the antenna) have as well being carried out using a differential technique showing great agreement with simulations [24]. For noise figure, several methods were used, namely differential Y-factor measurements using liquid Nitrogen, measurements with a noise figure analyzer as well as a noise tuner operating in the high end of the band. A differential Y-factor measurement was carried out to get an indication of the Johnson noise generated by a 150Ω chip resistor as the input load. This was chosen as it was close to the average differential impedance of the antenna across the band and thus signified the impedance seen by the LNA. Liquid Nitrogen was used as the cold source at 77K and room temperature (290K) represented the hot source. Each input of the board was fed with a short coax (RG402 - stainless steel) and the soldered chip resistor could then be dipped in the liquid Nitrogen dewar. This type of measurement is extremely sensitive and great care had to be taken to ensure that the LNA board itself did not freeze. A plot of the measured noise temperature is shown in Fig. 26. At higher frequencies (>400MHz), the noise temperature increased due to the length of the coax cable (~6cm) changing the input impedance. This result was then verified in the single-ended case (1 input loaded) using a noise figure analyzer (NFA). The LNA board was connected to the test system using a SMA connector at the active input. Figure 26 shows, together with the measured noise temperature, the simulated noise temperature when the differential amplifier is loaded with the antenna impedance (note that in the measurements the amplifier was loaded with a fixed impedance). The measured levels were considered satisfactory. The fluctuations seen in Fig. 26 are mainly a function of the NFA which has a measurement accuracy of ±0.1dB. Furthermore, we carried out a single-ended noise parameter measurement at ASTRON (Netherlands Institute for Radio Astronomy). This was achieved with the aid of a noise tuner whilst the balun and second stage amplifier were removed. However, since the tuner was rated to work typically from 400MHz up to GHz range, we were unable to achieve worthy results at frequencies lower than ~370MHz. With the input provided from these measurements, Fig. 27 shows a plot of the minimum noise temperature achievable with the LNA and that of the LNA loaded with the impedance of the antenna. It is clear that in this band an average noise temperature of ~35K is achievable. The input IP2 and IP3 of the LNA were determined using a 2-tone experiment for both, the single amplifier and the pseudo-differential configuration with the second stage gain, delivering similar results in both of -23dB and -18dB for IP2 and IP3, respectively. These represent the worst recorded cases.

### C. Array measurements

Two major prototype arrays have been built to date using SKALA elements. These arrays have been built by the institutions forming the LFAA consortium led by ASTRON, Netherlands (this is the consortium in charge of developing the LFAA instrument). The first one was AAVS0 (Aperture Array Verification Systems 0) and following this, LFAA built AAVS05. AAVS0 was built at Lords Bridge observatory, Cambridge, UK and AAVS05 at the MRO (Murchison Radio-astronomy Observatory) site (future location of the SKA-low telescope). Both arrays have 16 antennas, which in the case of AAVS0 can be deployed in any desired configuration. We have tested AAVS0 in regular as well as irregular configurations. These arrays have been extensively tested to validate the designs and simulations of the front-end (antenna and LNA) specifically. Tests of AAVS0 range from impedance and scattering parameters tests [28] to near field pattern tests using a known source [29]. Other tests include measurements of the cross-polarization capabilities of the antennas [30]. The AAVS05 array, working in conjunction with the MWA telescope, has been able to produce test results using astronomical sources, such as a drift-scan of the Galaxy [31]. The AAVS05 array, in contrast to AAVS0, does not have a ground screen under the antenna elements, which is being useful to analyze the effect of the ground on the antenna performance (noise temperature, pattern, etc.). Figure 28 (a) shows an image of AAVS0 and 28 (b) an image of AAVS05. Figure 29 shows the coupling (measured and simulated) between 2 adjacent elements in the AAVS0 array. This confirms that the simulation tools used for the design of the antenna in the array environment work and can be used to save time in the development process.

AAVS0 is currently being upgraded with a digital back-end to continue and expand the tests for the next version of SKALA. These tests will include interferometry tests using astronomical sources as well as artificial sources. The plan of the LFAA consortium is to build a 3-station demonstrator (AAVS1) with 256+64+64 SKALA elements in the MRO site by the end of 2015. This demonstrator will be used for tests and validation of the LFAA technology being currently developed, including not only the front-end but also the receiver, communications, signal processing and infrastructure.

## 6. Conclusions

In this paper we have presented the design of an active element for the SKA-low instrument. An innovative variation of a LPDA element, composed of 9 dipoles, was designed keeping in mind the main requirements for SKA-low and the main constraints. Good cross polarization, a small foot-print and maximum sensitivity across the field of view were three of the main target parameters. Moderately high directivities offered by LPDA elements have been used to optimize the performance for the specific application of SKA.

We have described the main design targets and implications on the overall system and presented simulated results as well as validation of the simulations with measurements. This validation of the simulated designs means that we can proceed with high confidence the development of SKALA until it is finally ready for deployment for Phase I of SKA by 2017. We are currently going through this development phase where the mechanics of the antenna (for an expected life time of 30 years+) and the integration with the rest of the LFAA system are a key part of the work. This evolution of the design, will try to achieve a *ready for deployment* state where 250,000 SKALA elements will be ready for mass production and easy deployment while meeting the demanding specifications of the SKA telescope.

## 7. Acknowledgments

The authors would like to acknowledge their SKA colleagues for numerous discussions and constructive input about the topic.

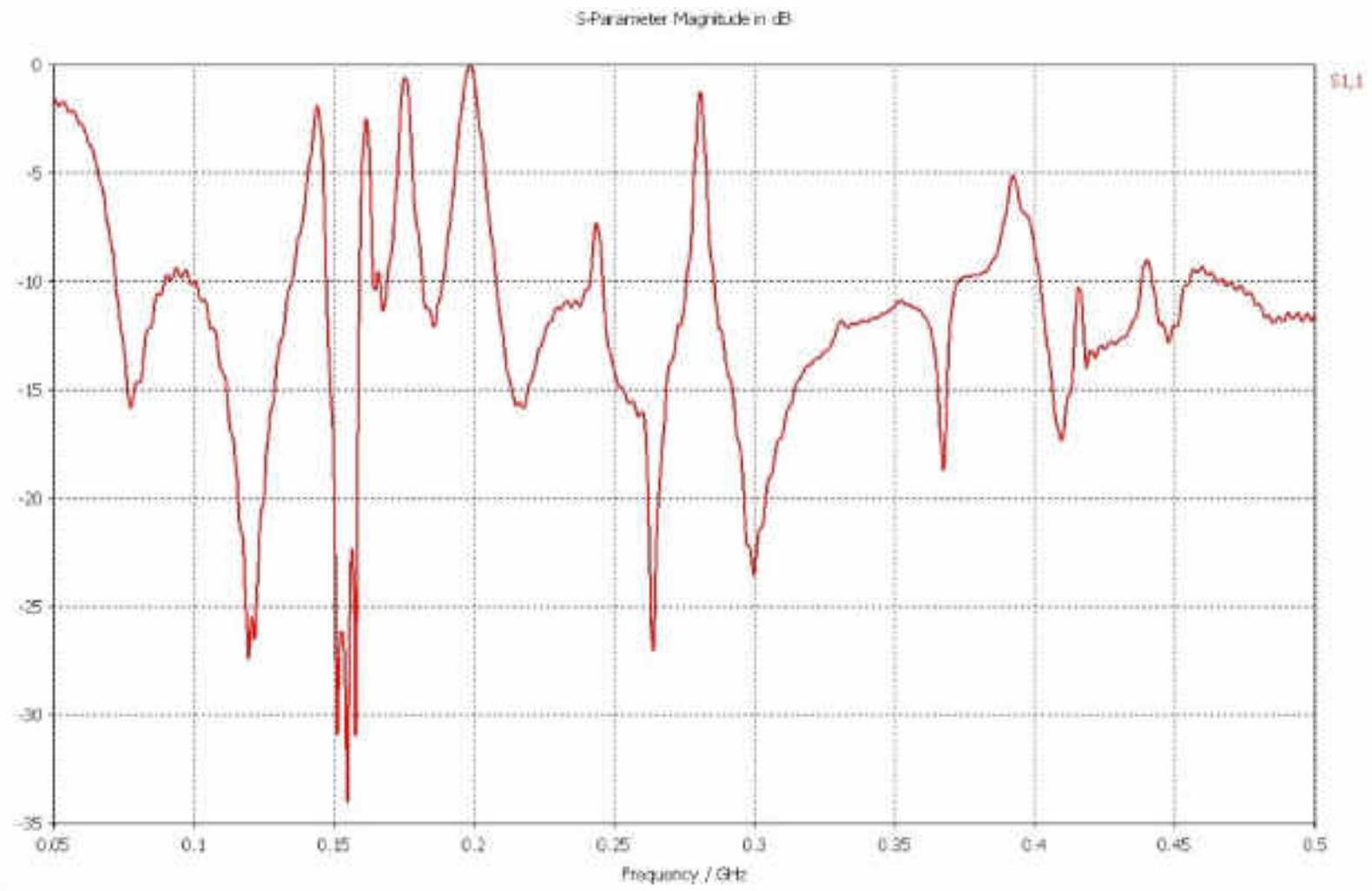

Fig. 1. Effects of mutual coupling in the active antenna element’s $S_{11}$ for a regular configuration. Element spacing is λ/2 at 100 MHz.

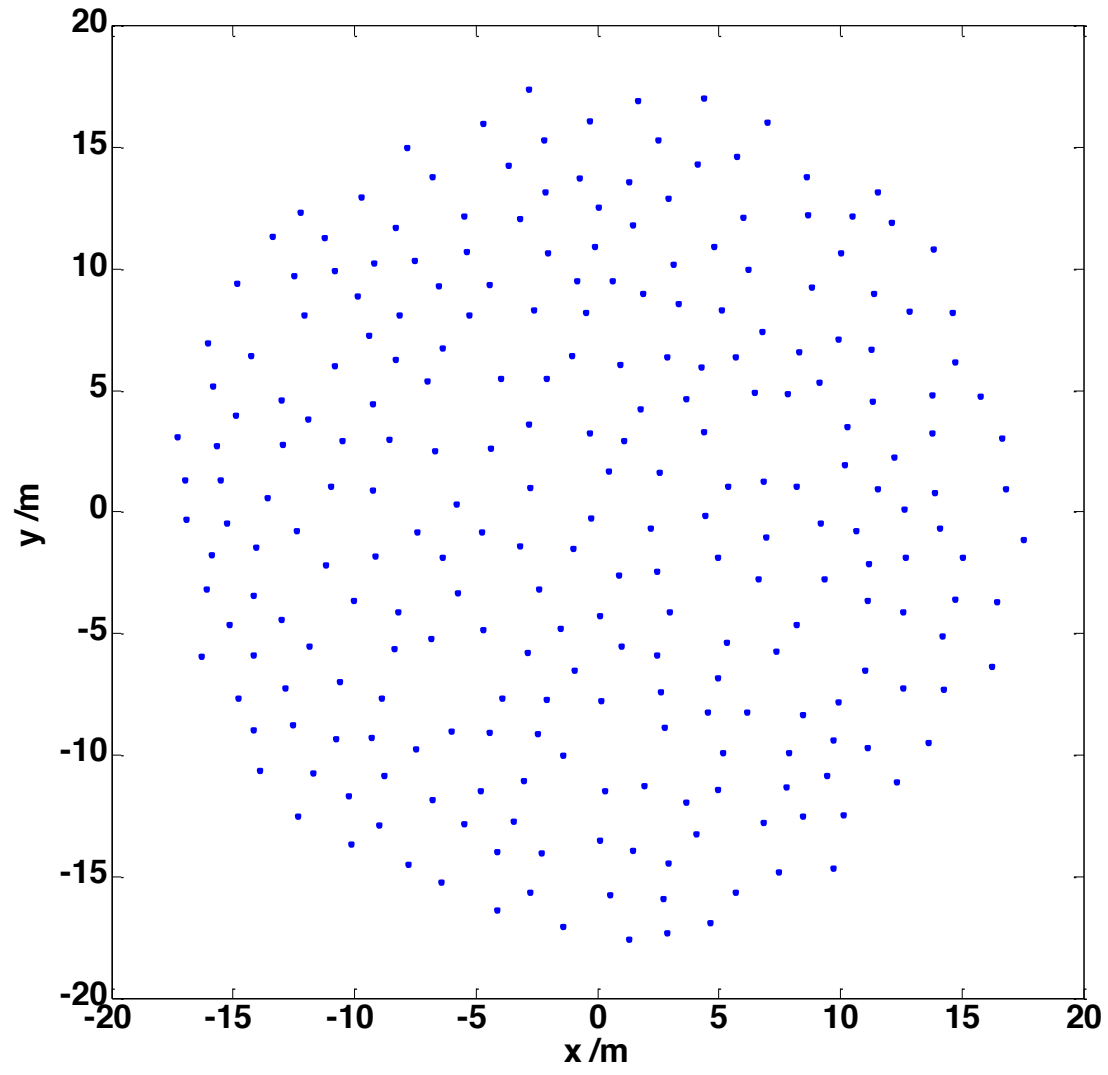

Fig. 2. Example of a 35 m station and 256 antenna elements with a random configuration.

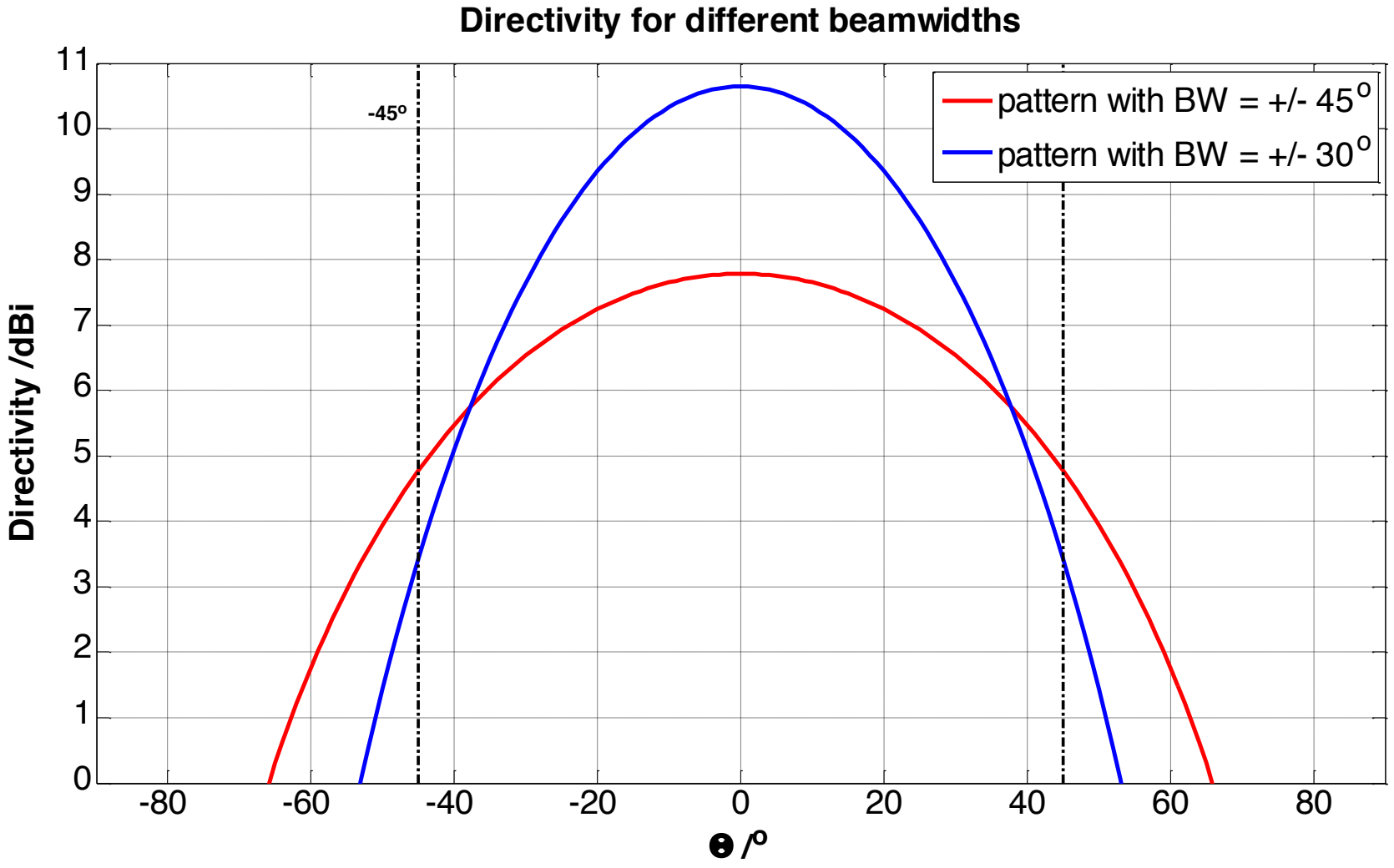

Fig. 3. Directivity comparison for beams with half power beamwidth of +/- 45º and 30º.

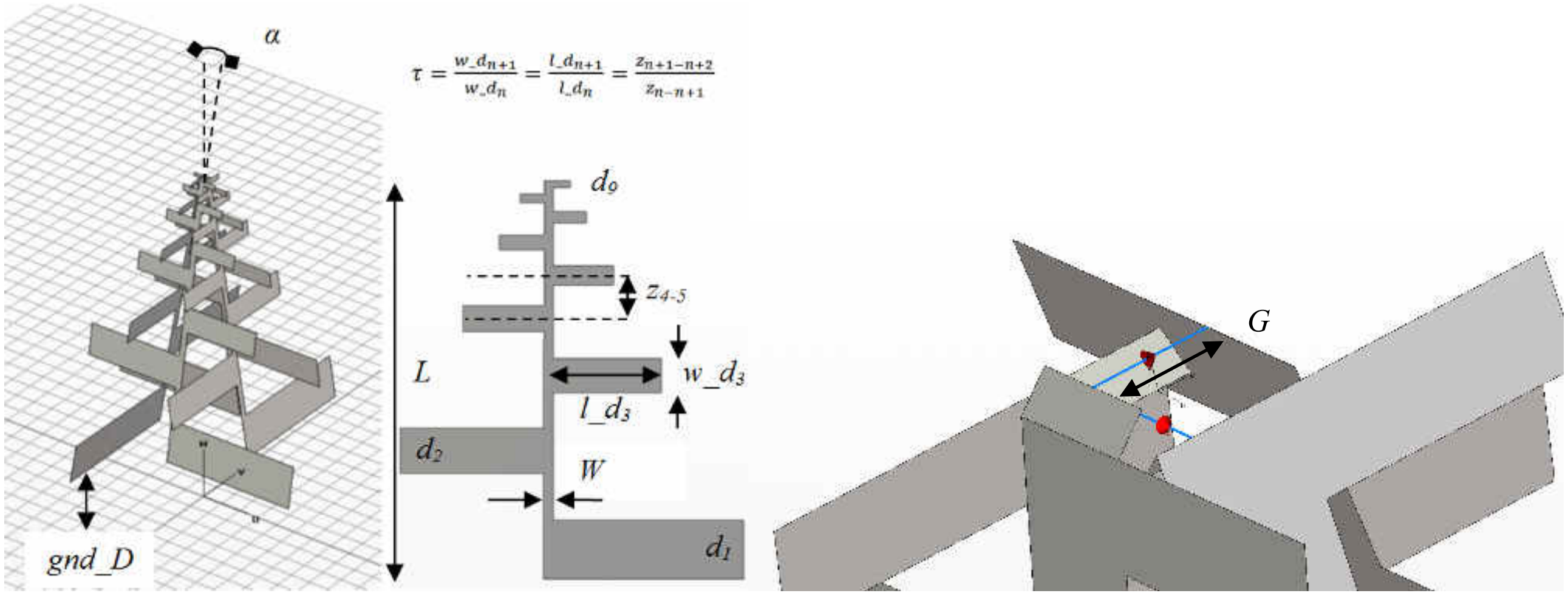

Fig. 4. Basic SKALA design (left), SKALA arm (center) and SKALA feed in simulations with 2 orthogonal non-overlapping ports (right), each of them featuring a 100 Ω series resistor with the generator.

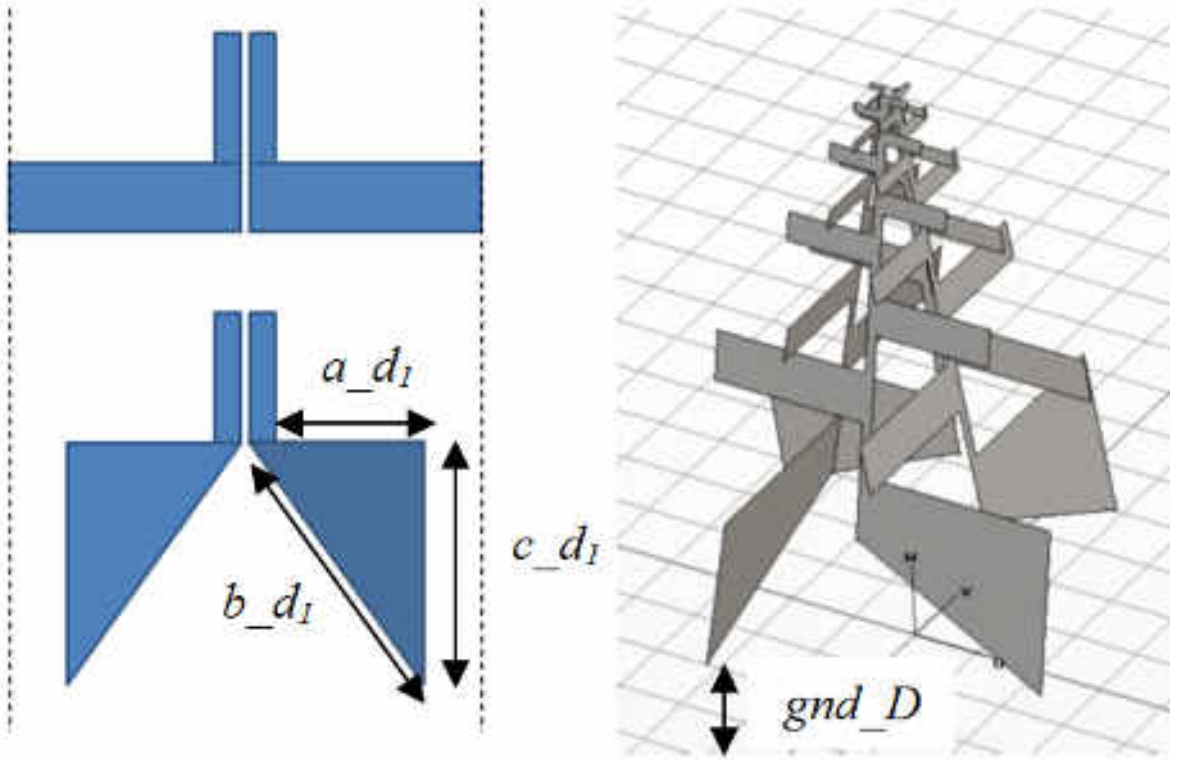

Fig. 5.  Modified bottom arm (left) and full modified design (right).

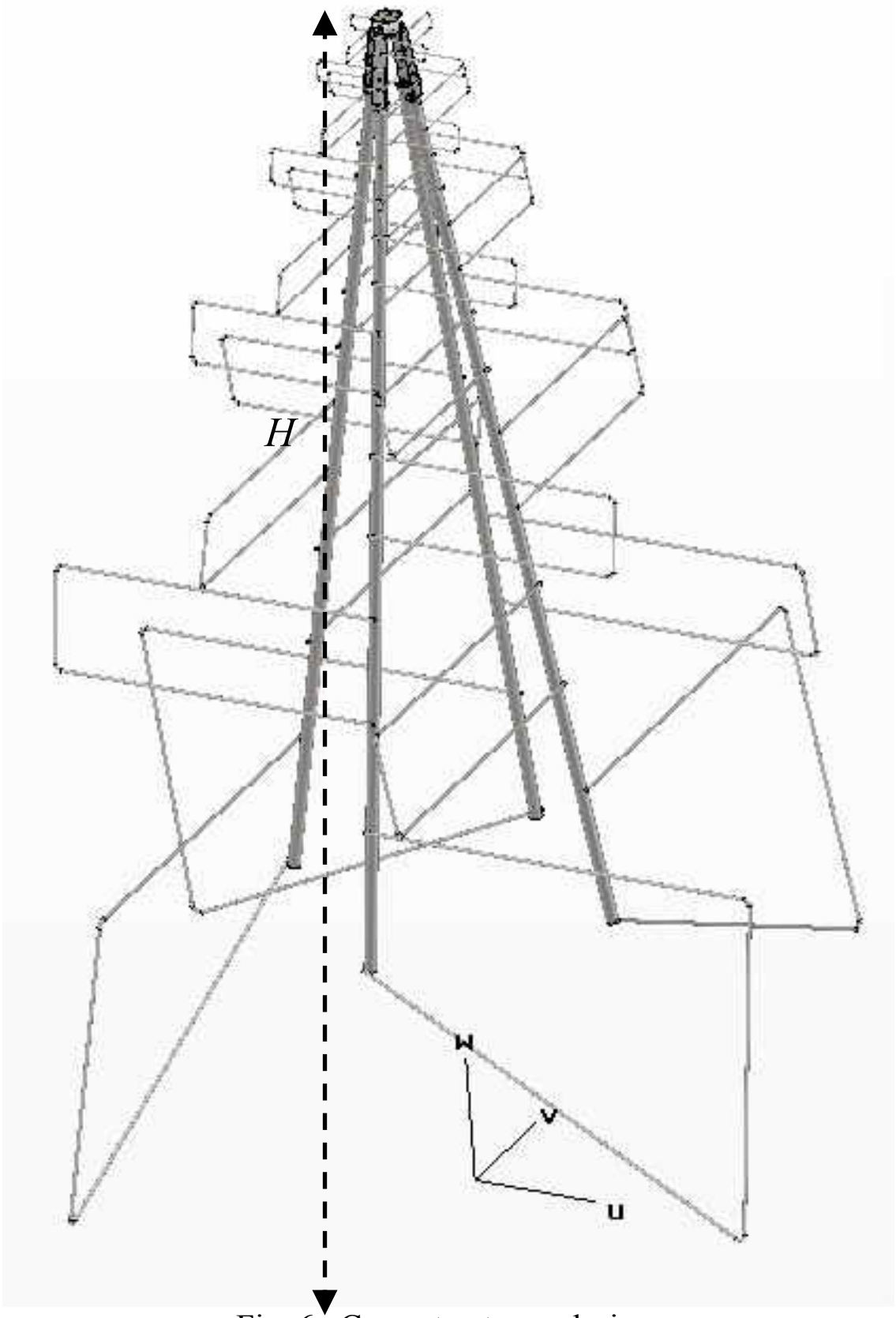

Fig. 6.  Current antenna design.

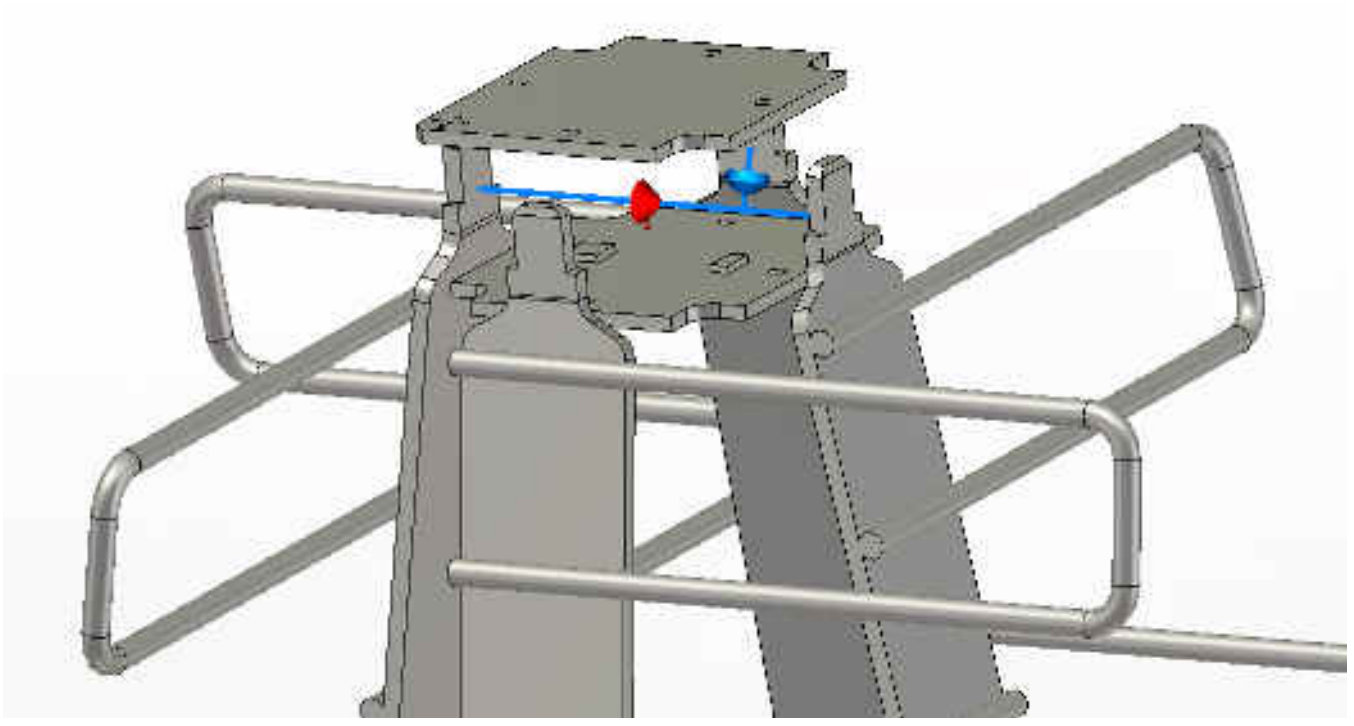
Fig. 7.  Feeding of current antenna including the LNA boards.

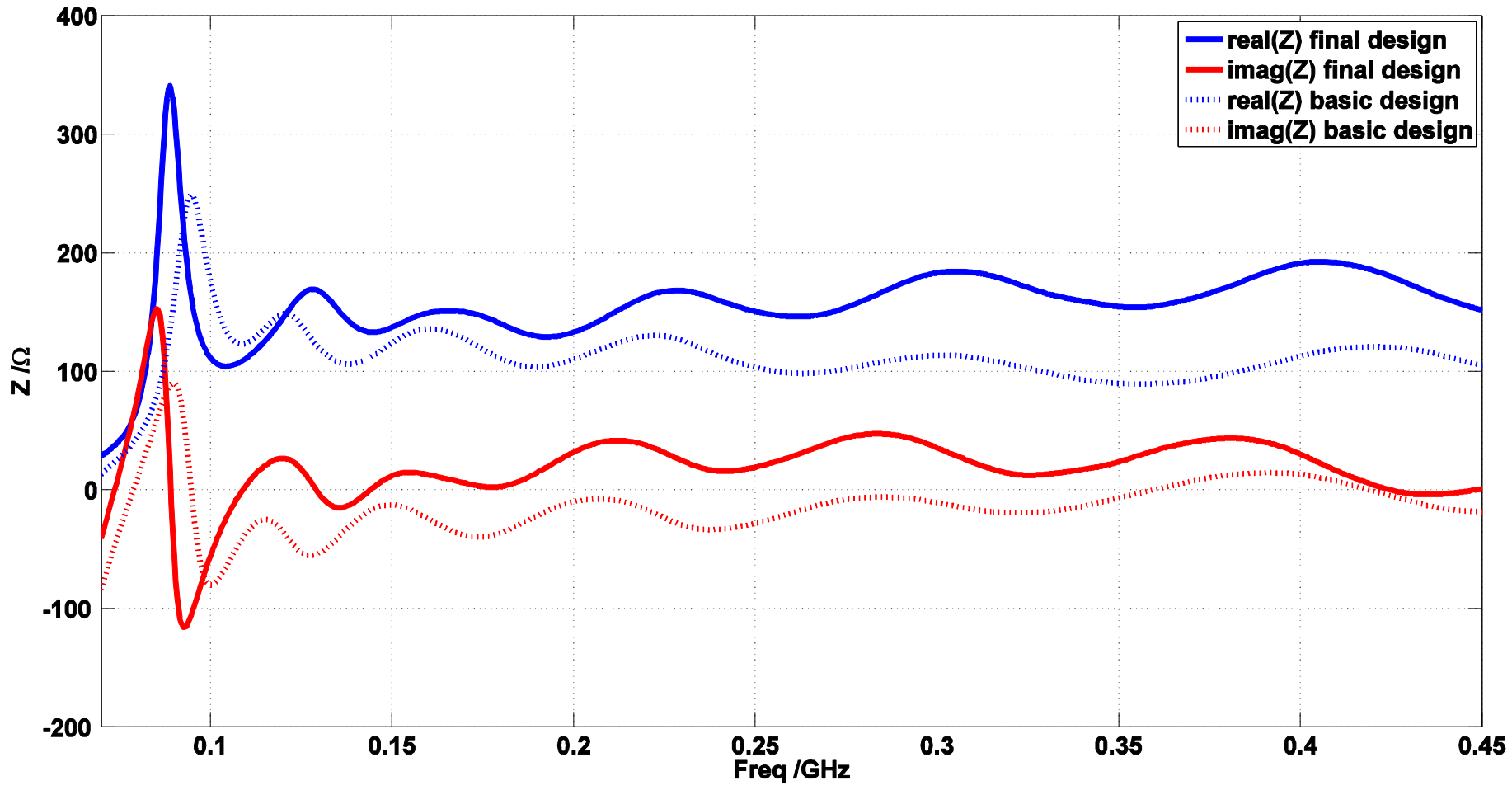

Fig. 8. Impedance of SKALA.

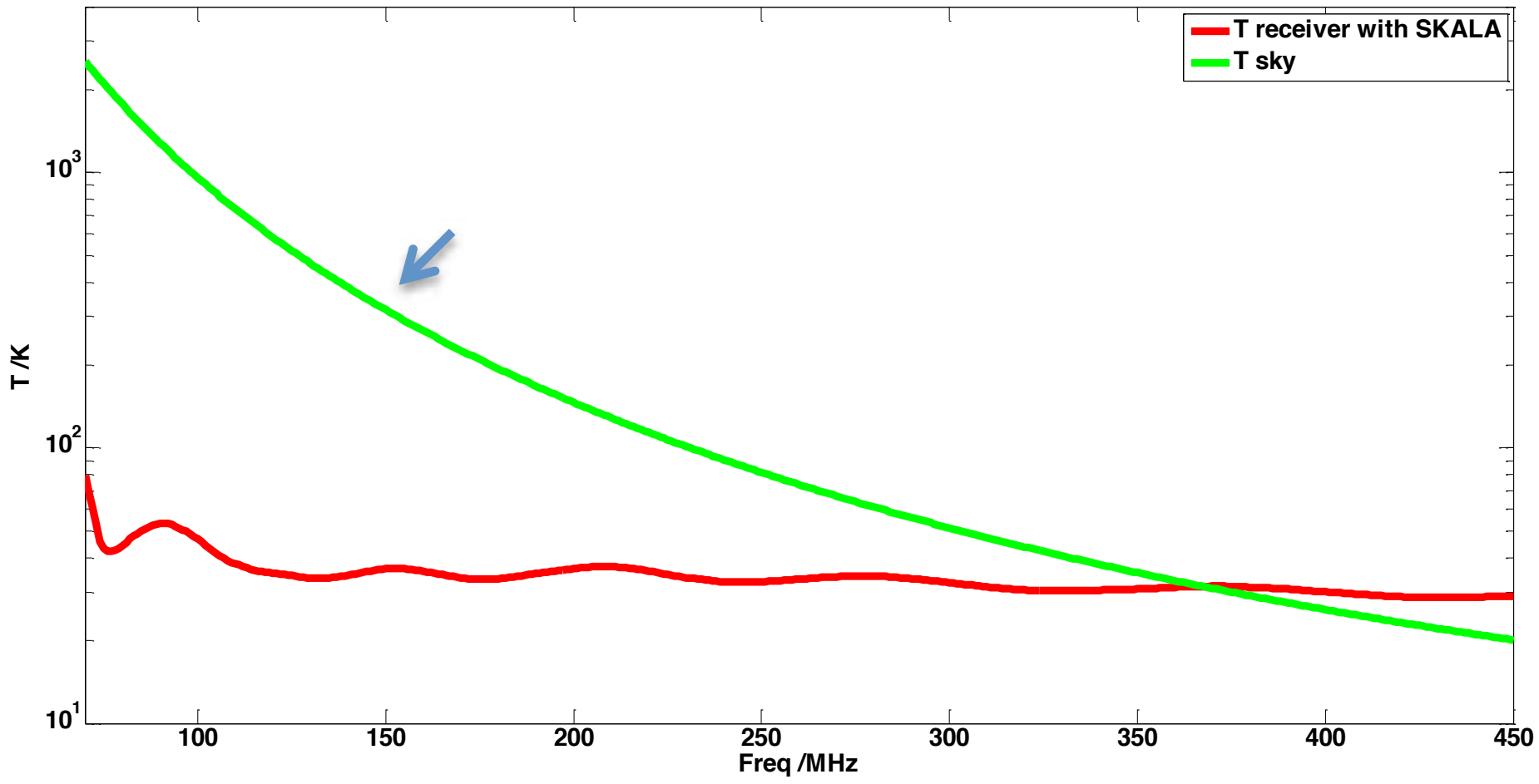

Fig. 9. Receiver noise temperature versus sky noise.

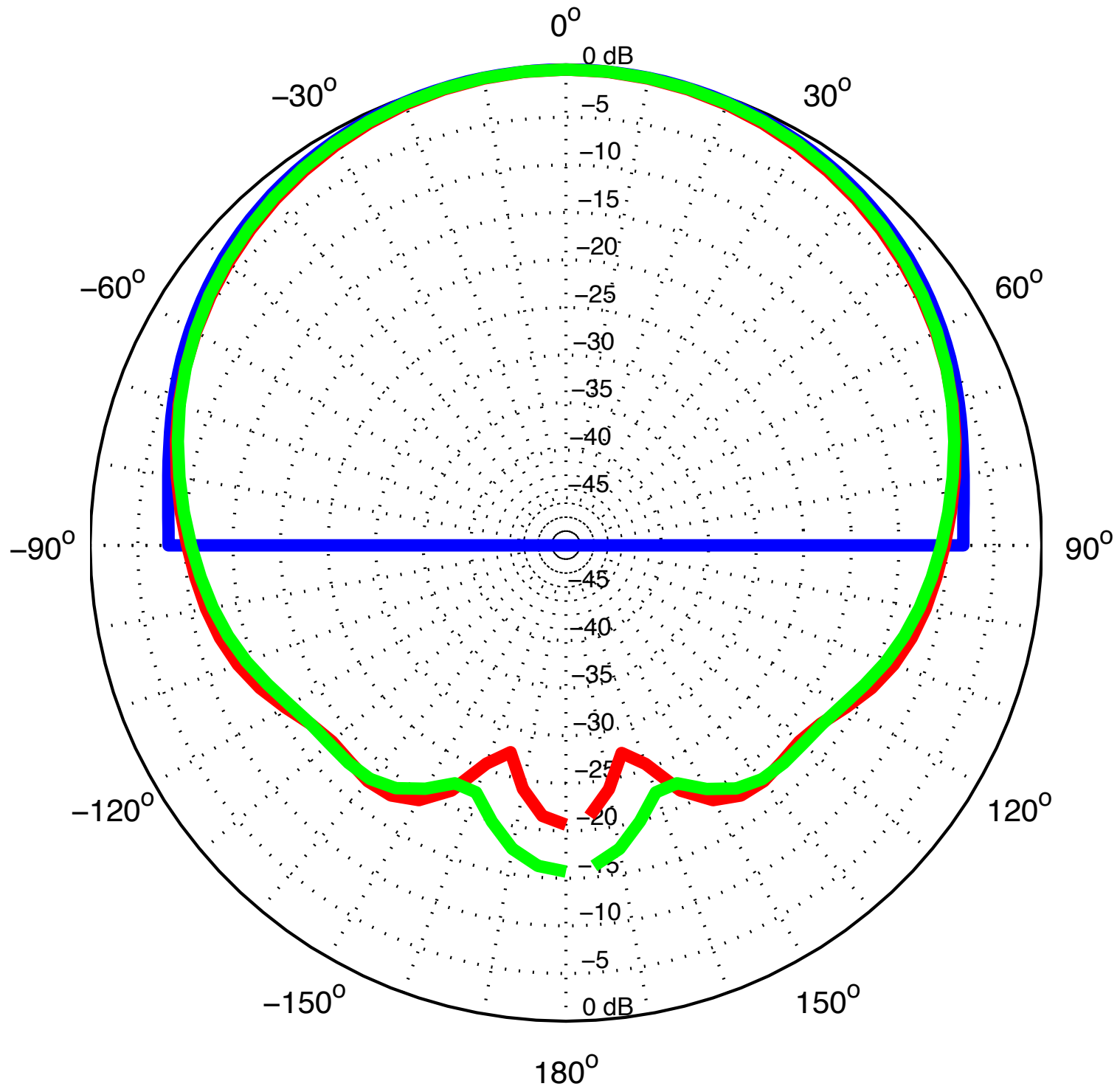

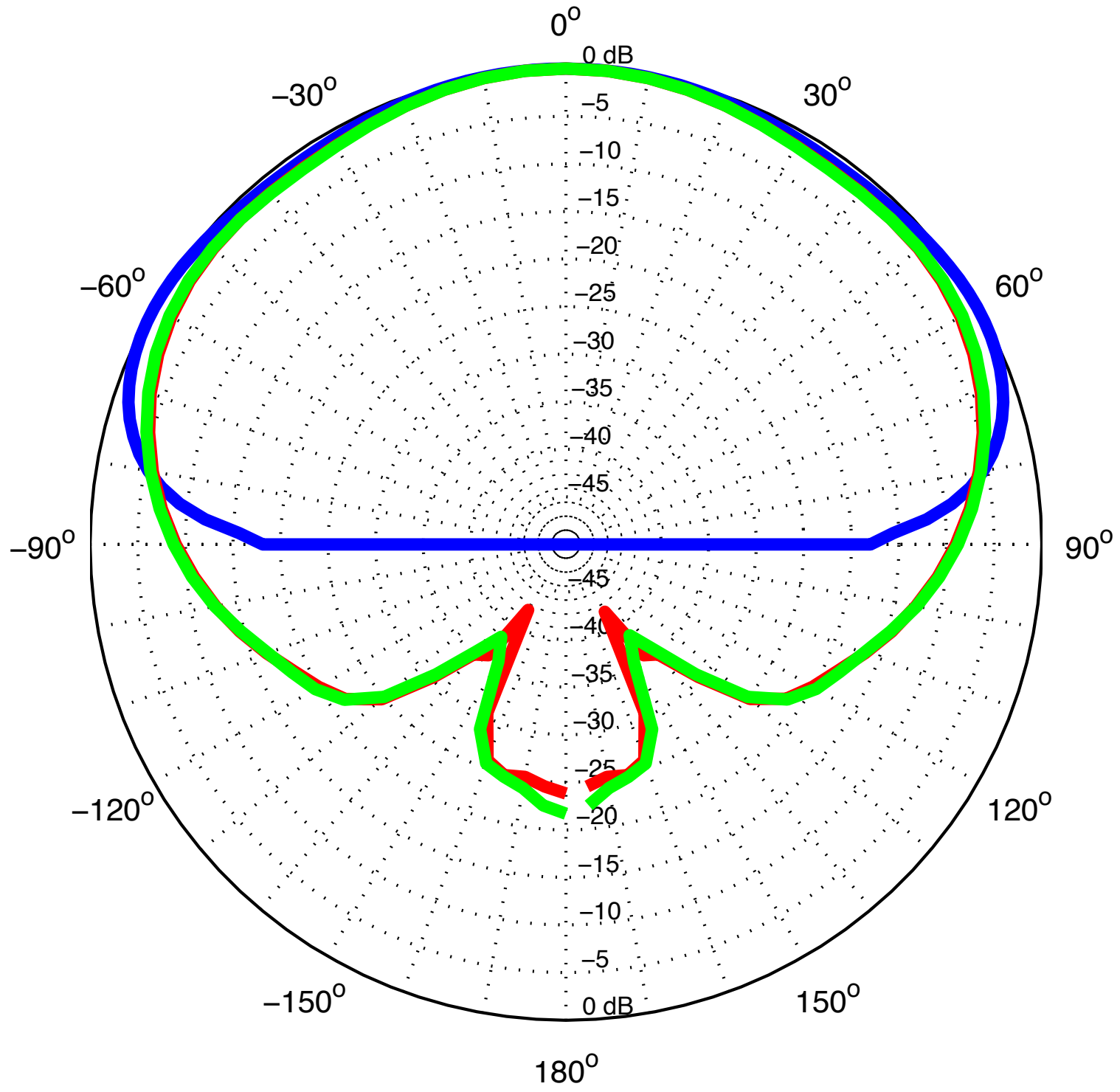

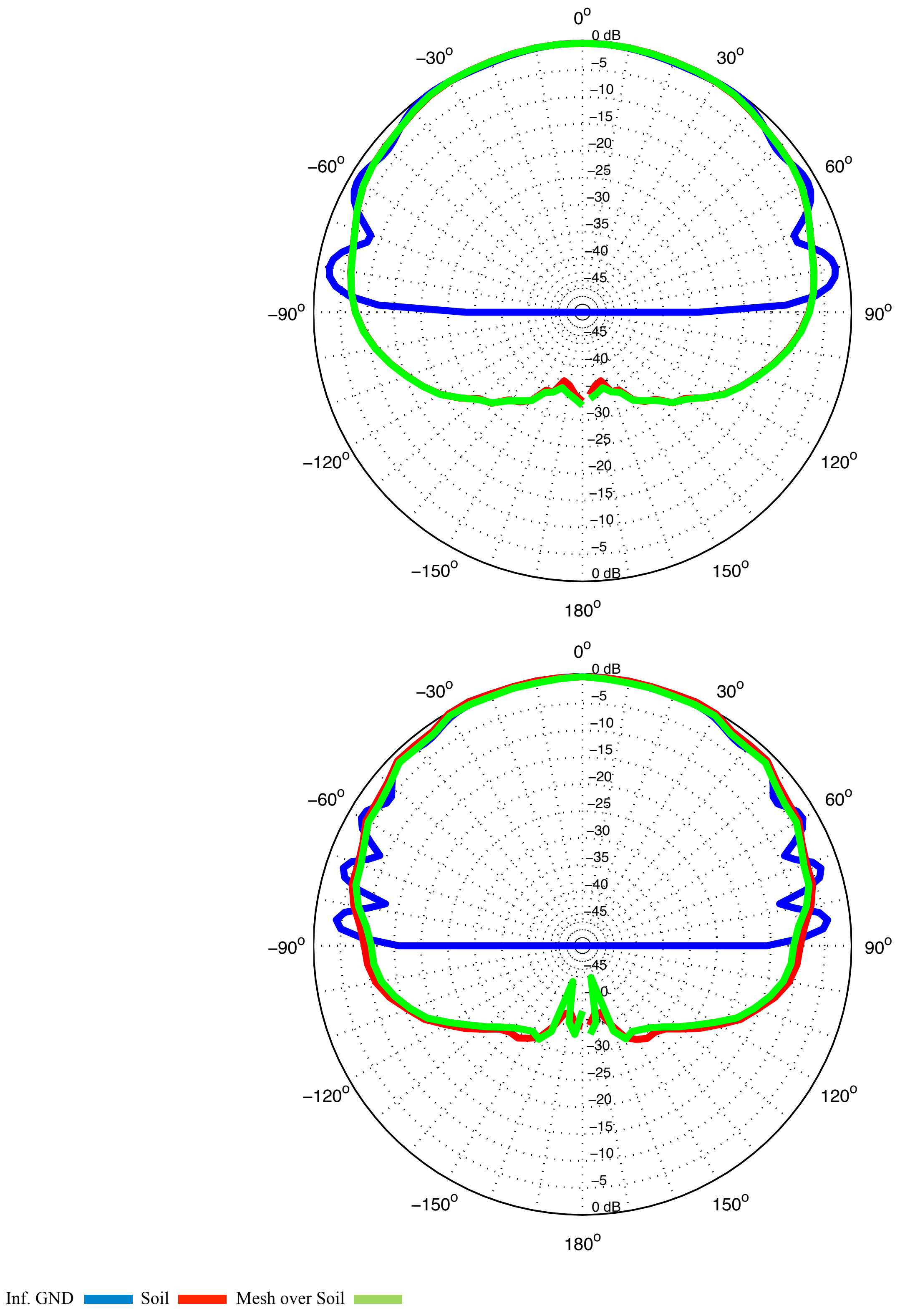

Fig. 10. Directivity pattern of SKALA (H-plane cut) at 70 MHz (top), 150 MHz, 300 MHz and 450 MHz (bottom).

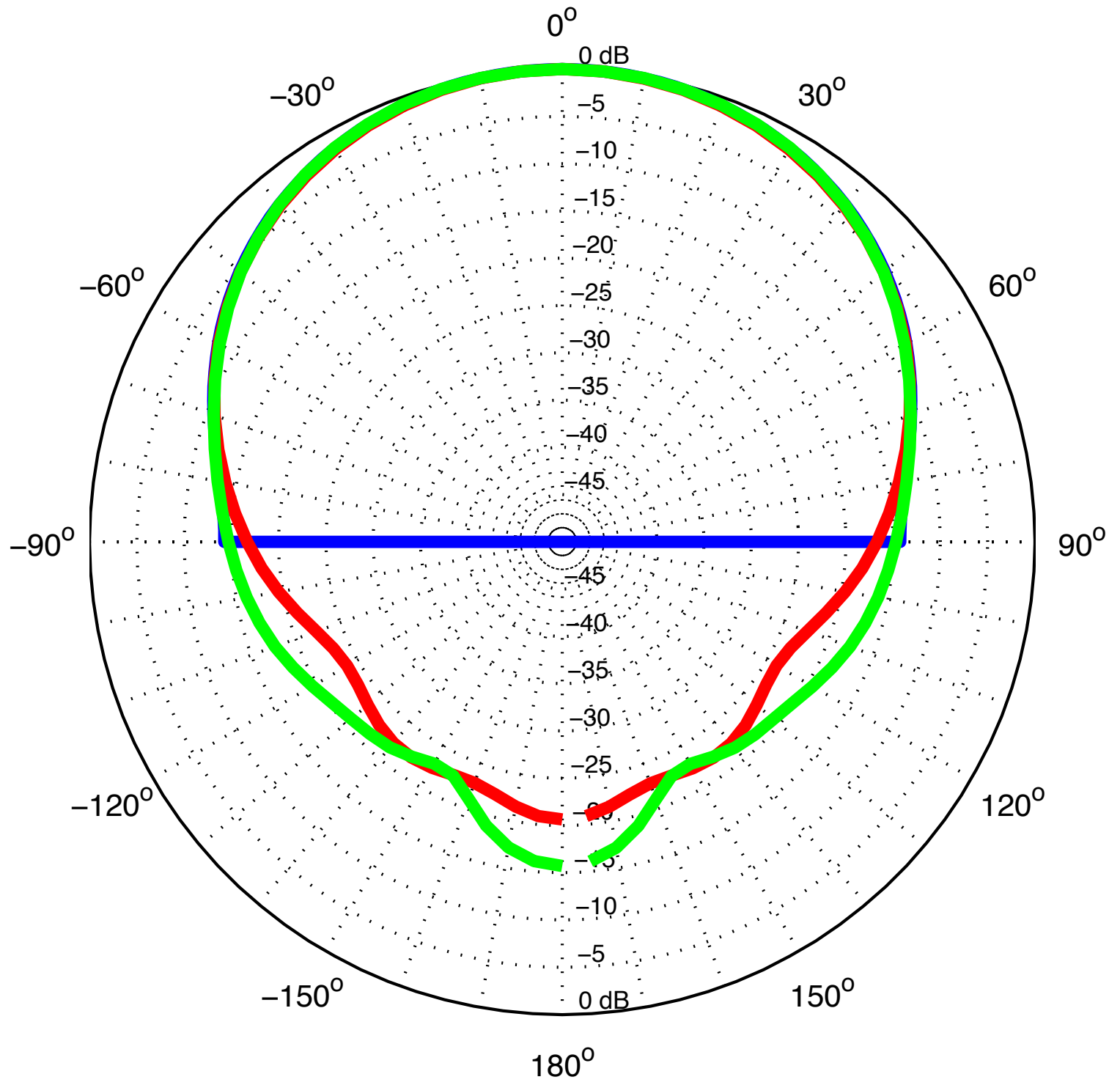

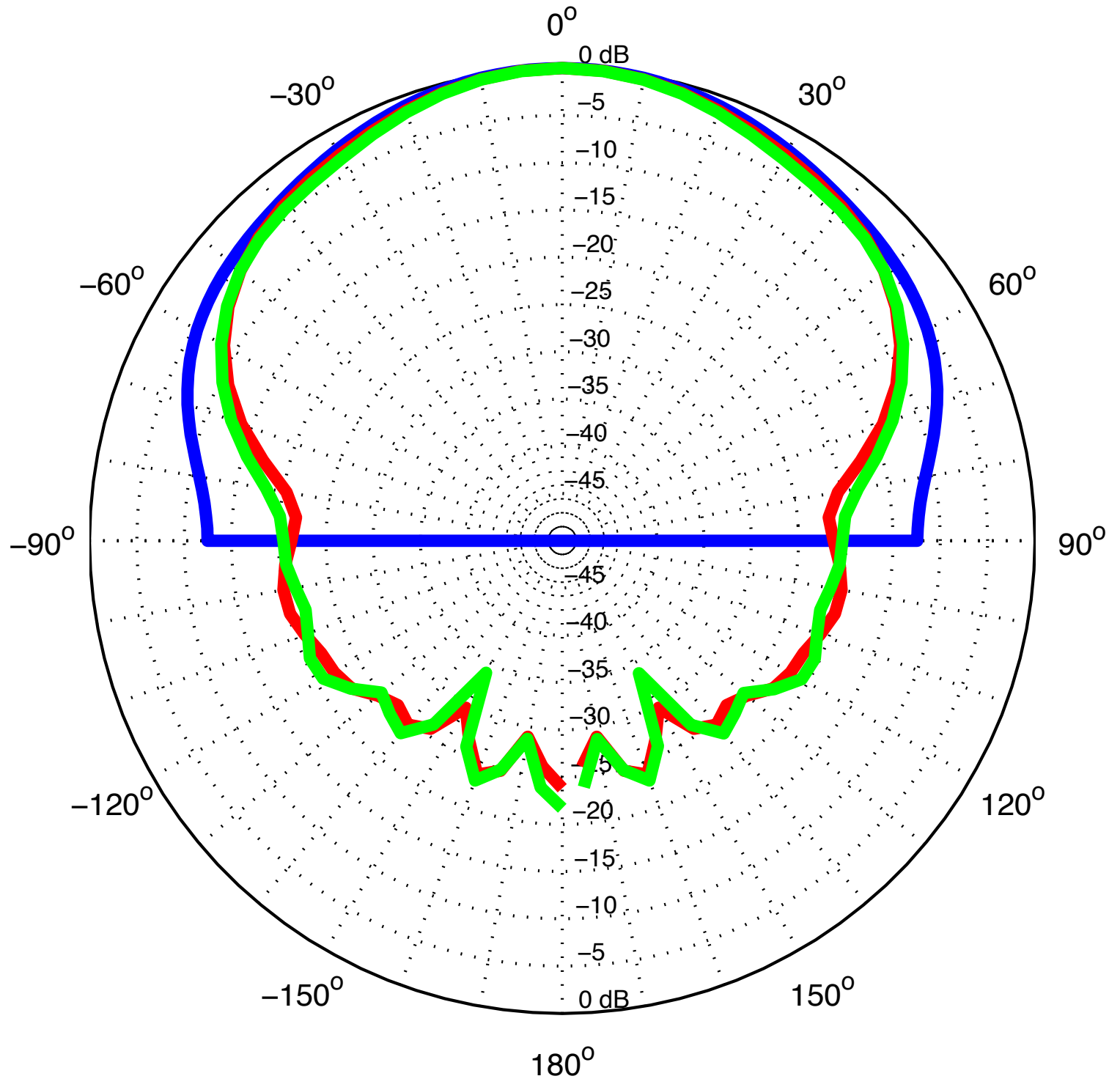

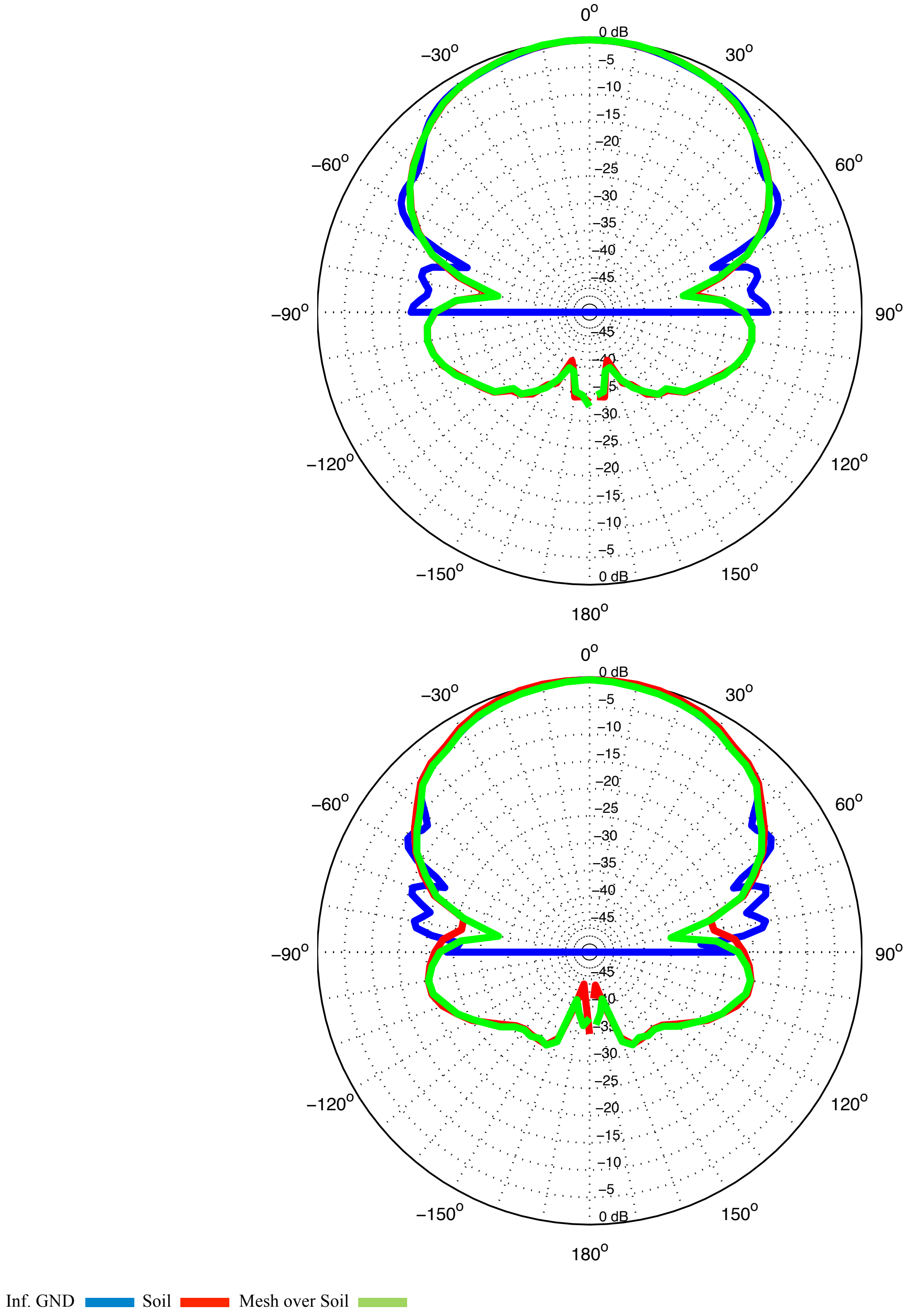

Fig. 11. Directivity pattern of SKALA (E-plane cut) at 70 MHz (top), 150 MHz, 300 MHz and 450 MHz (bottom).

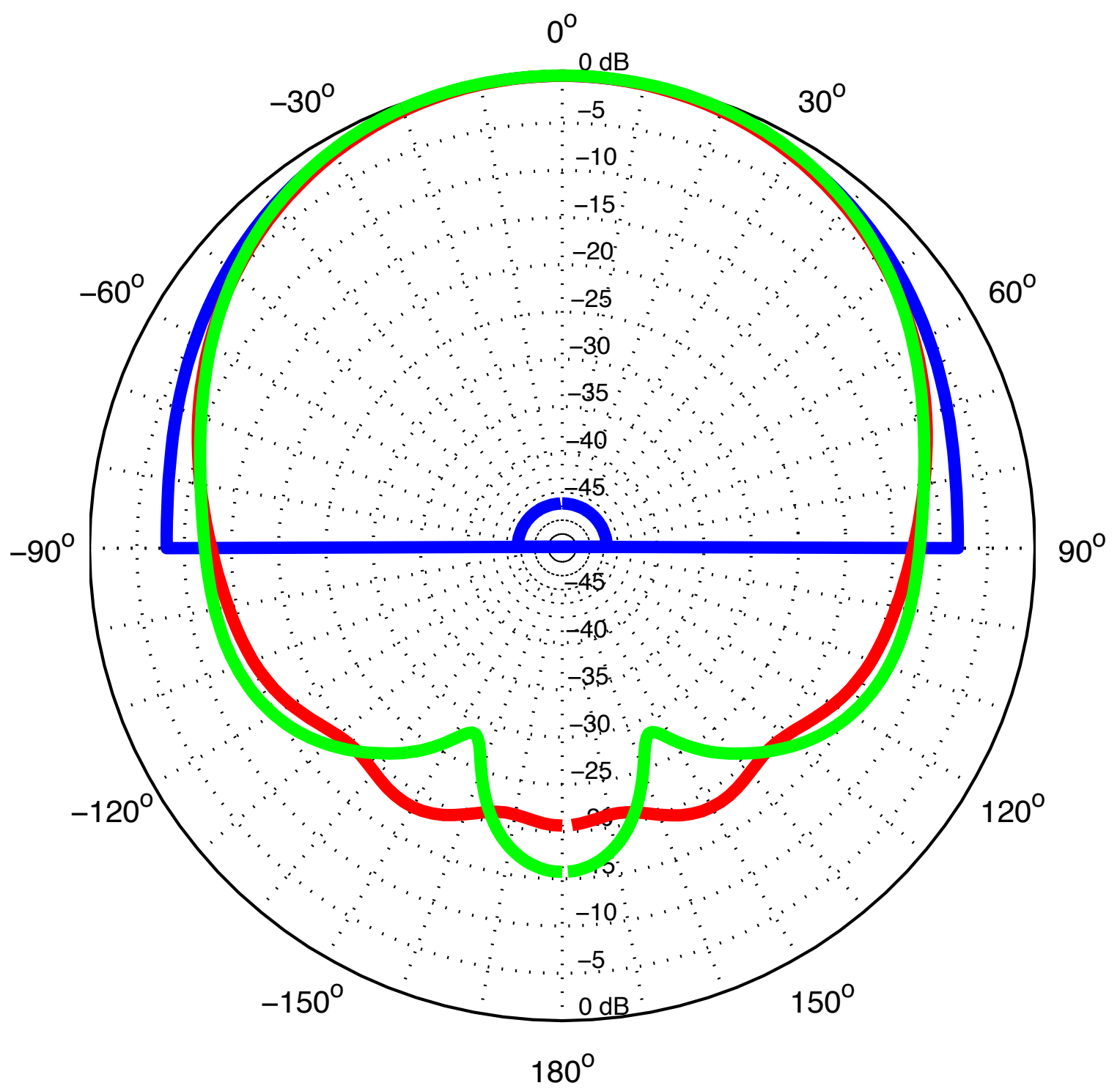

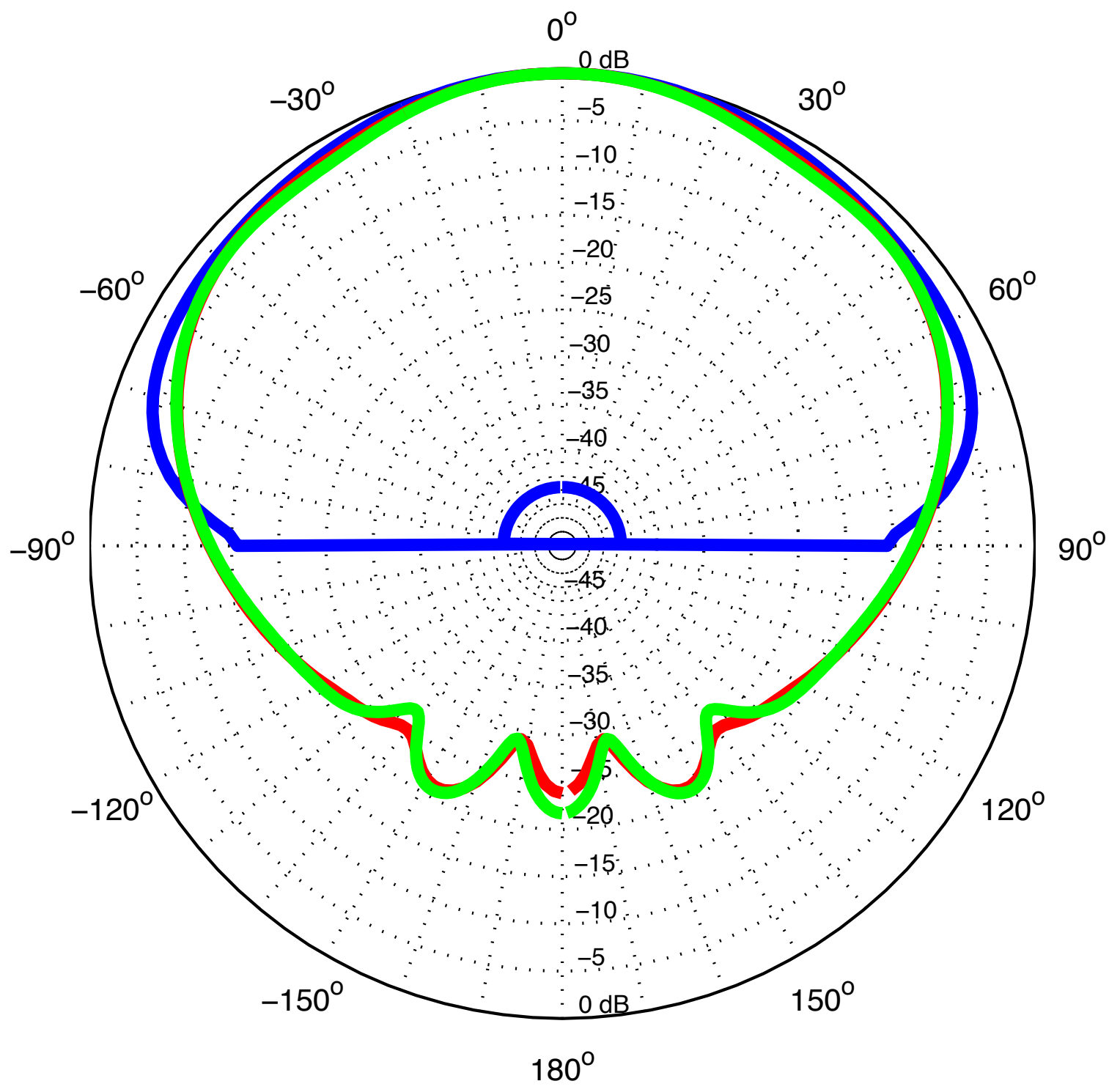

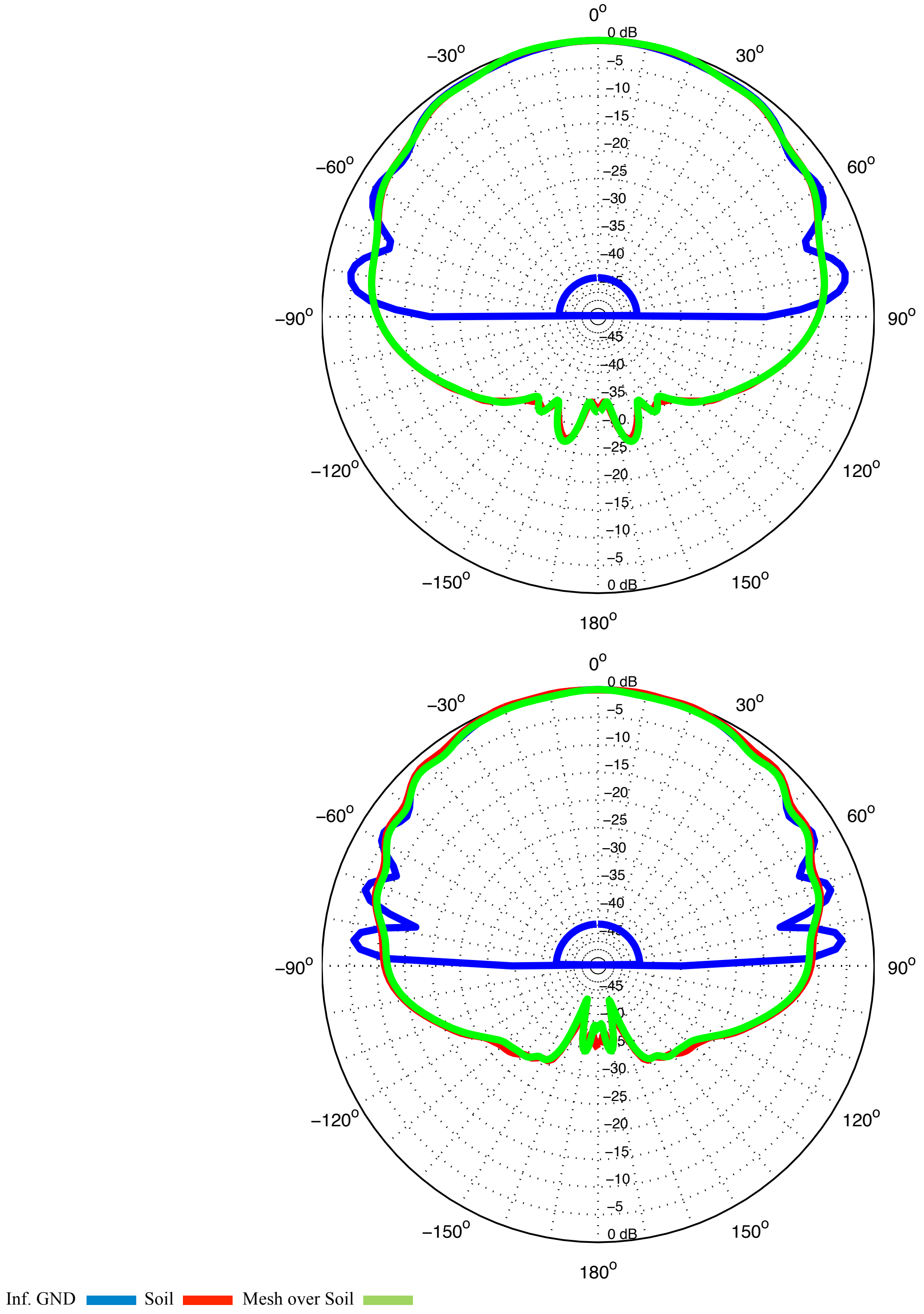

Fig. 12. Directivity pattern of SKALA (45° cut) at 70 MHz (top), 150 MHz, 300 MHz and 450 MHz (bottom).

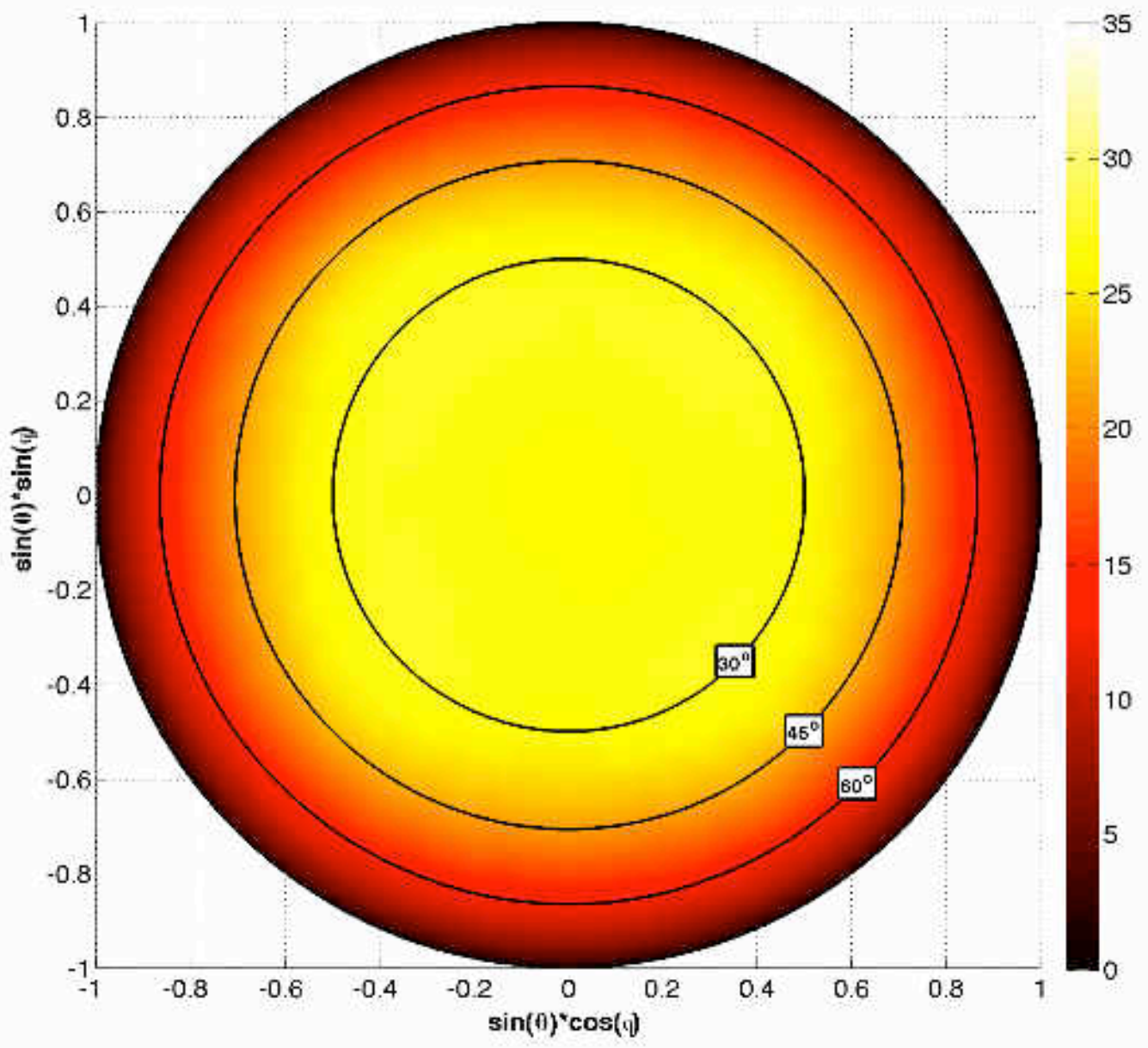
sin(θ)*sin(ψ)
sin(θ)*cos(ψ)
30°
45°
60°

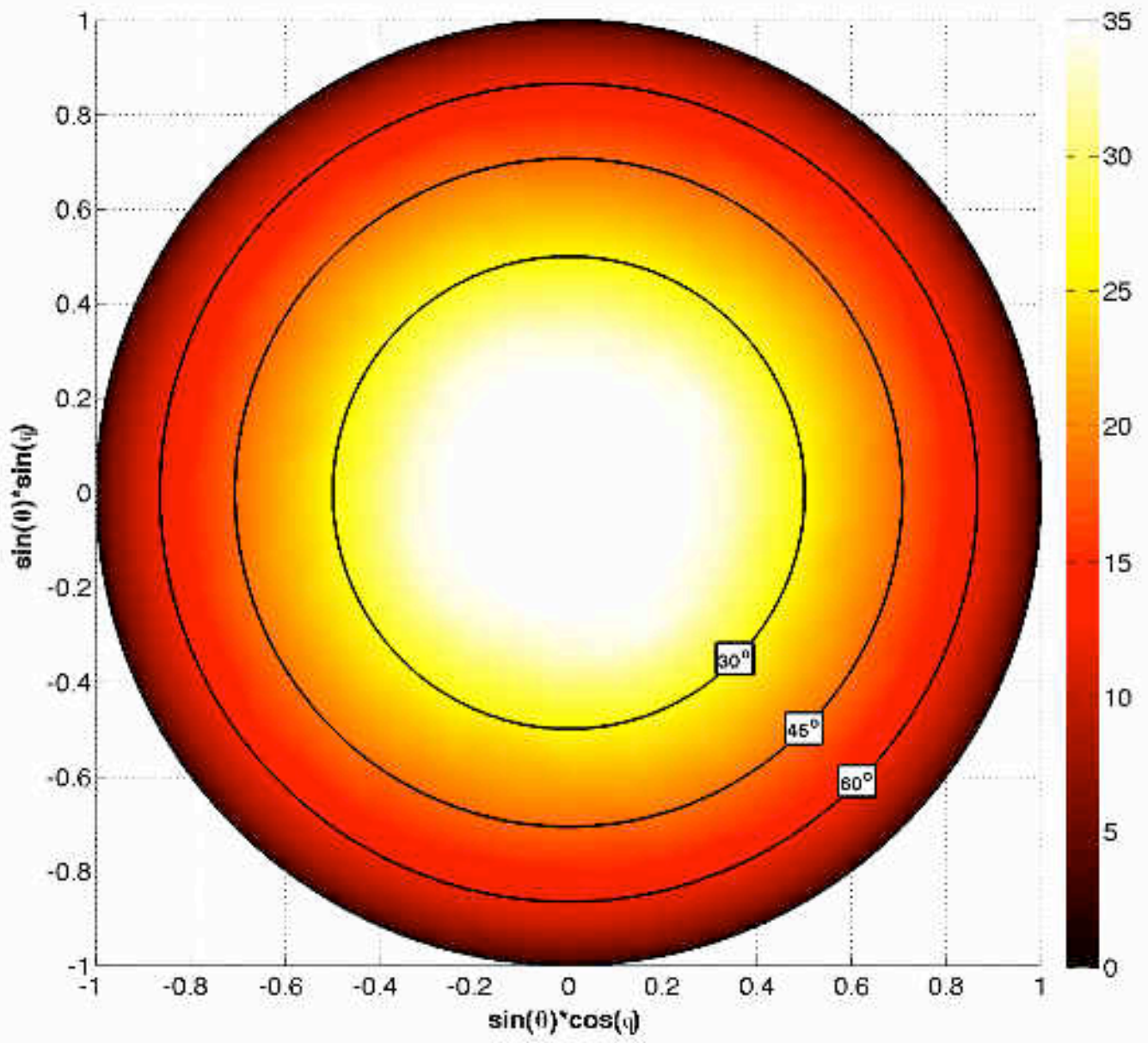
sin(θ)*sin(ψ)
sin(θ)*cos(ψ)
30°
45°
60°

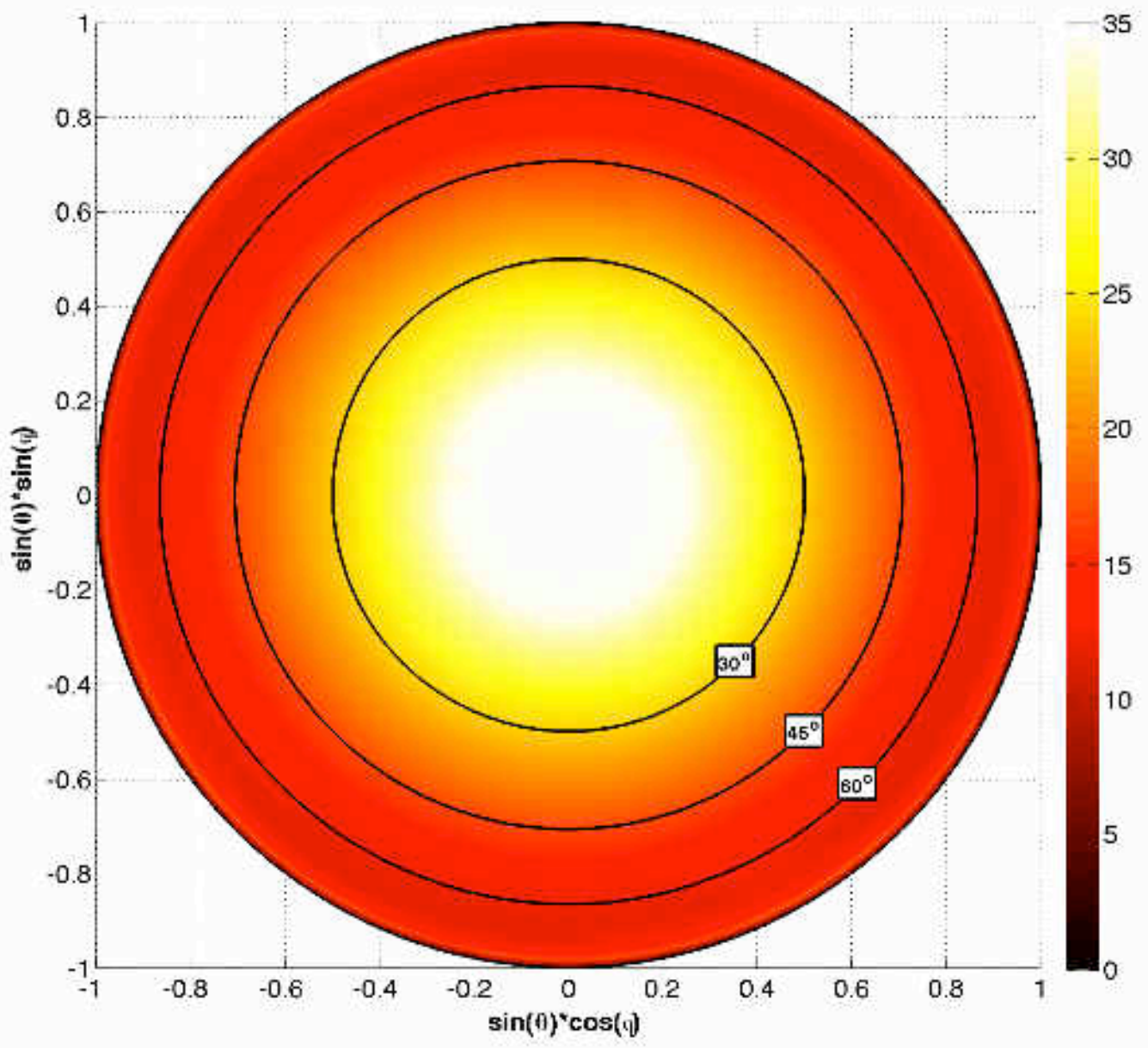

sin(θ)*sin(φ)
sin(θ)*cos(φ)
30°
45°
60°

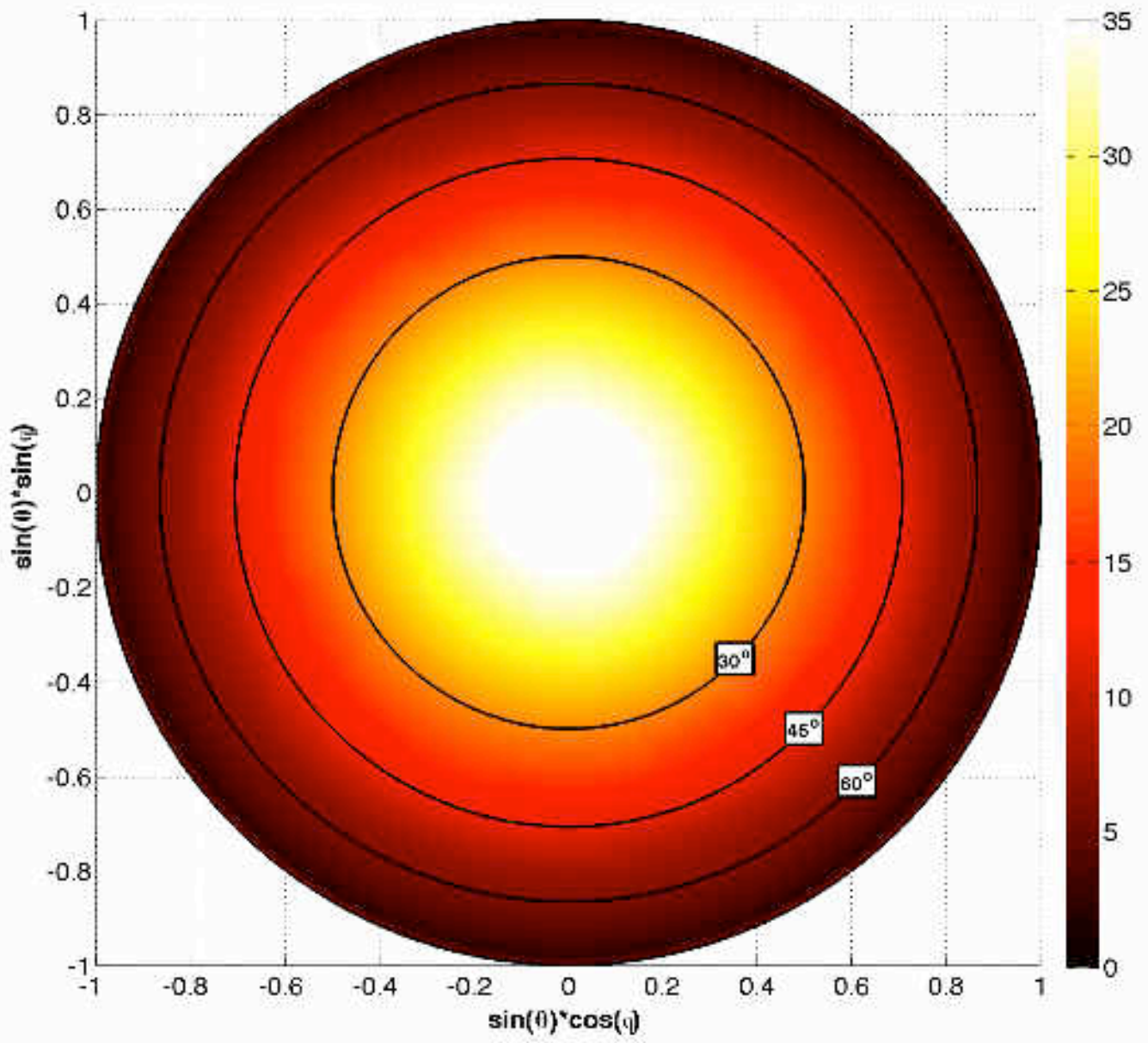

sin(θ)*sin(φ)
sin(θ)*cos(φ)
30°
45°
60°

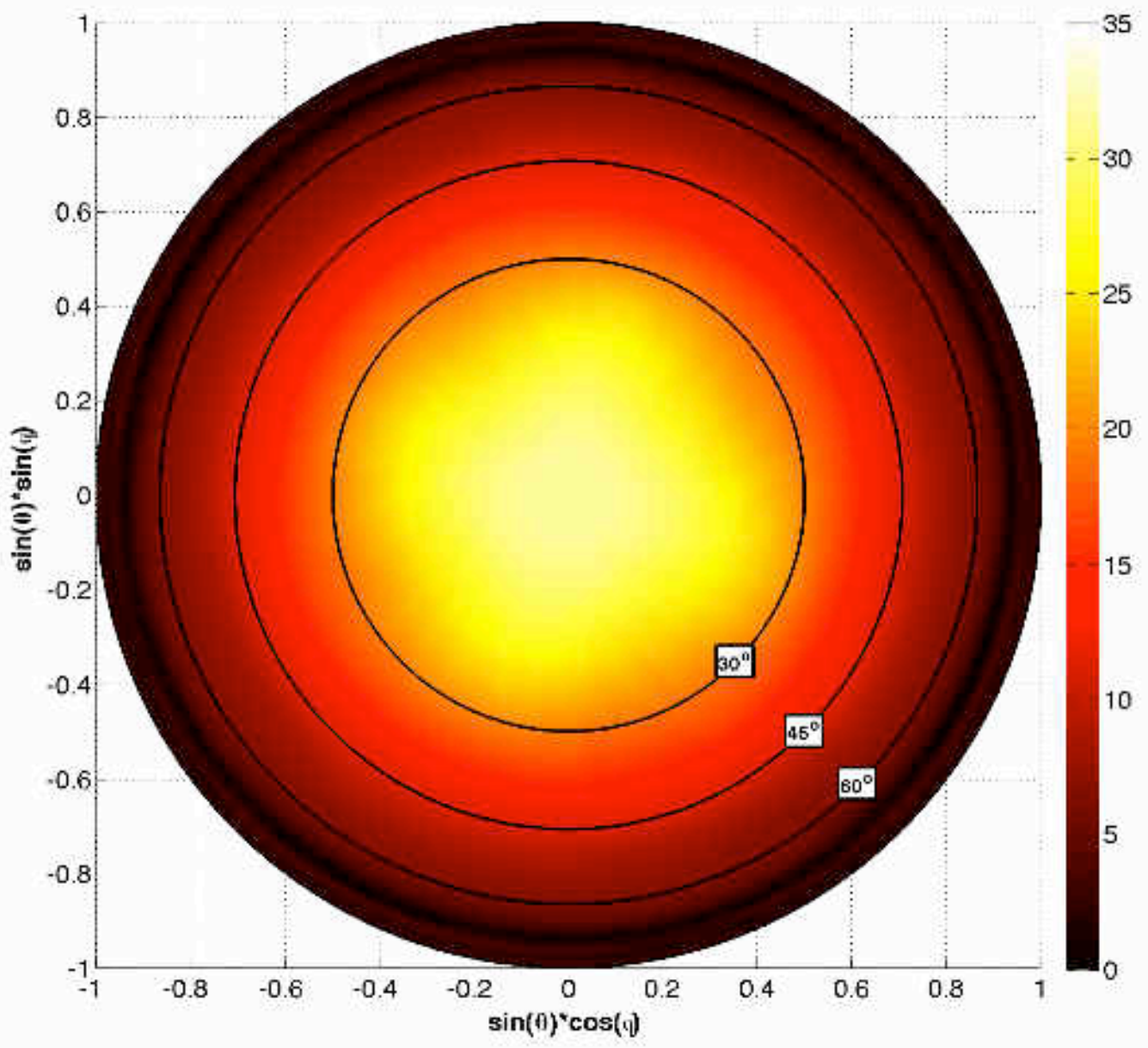

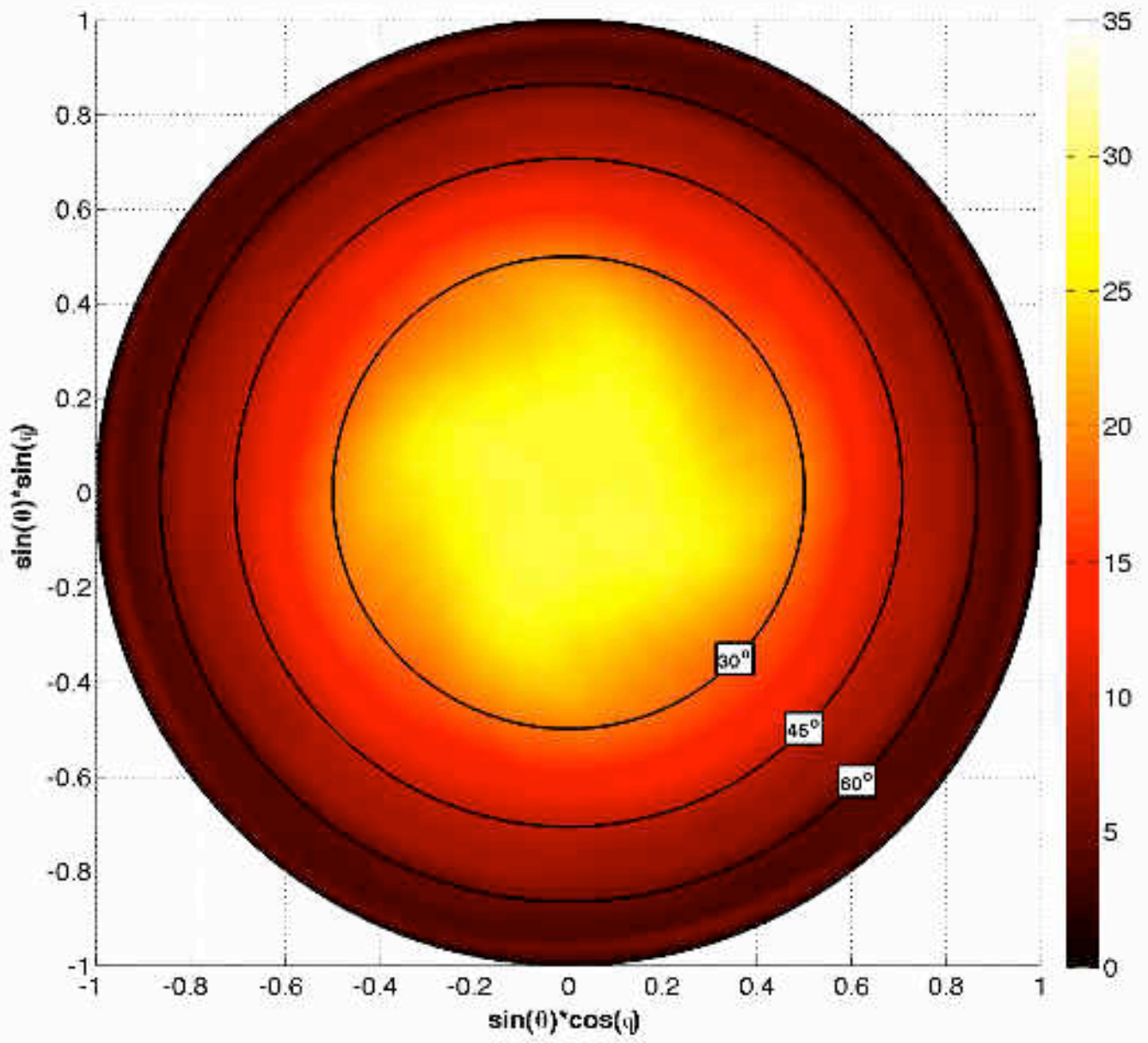

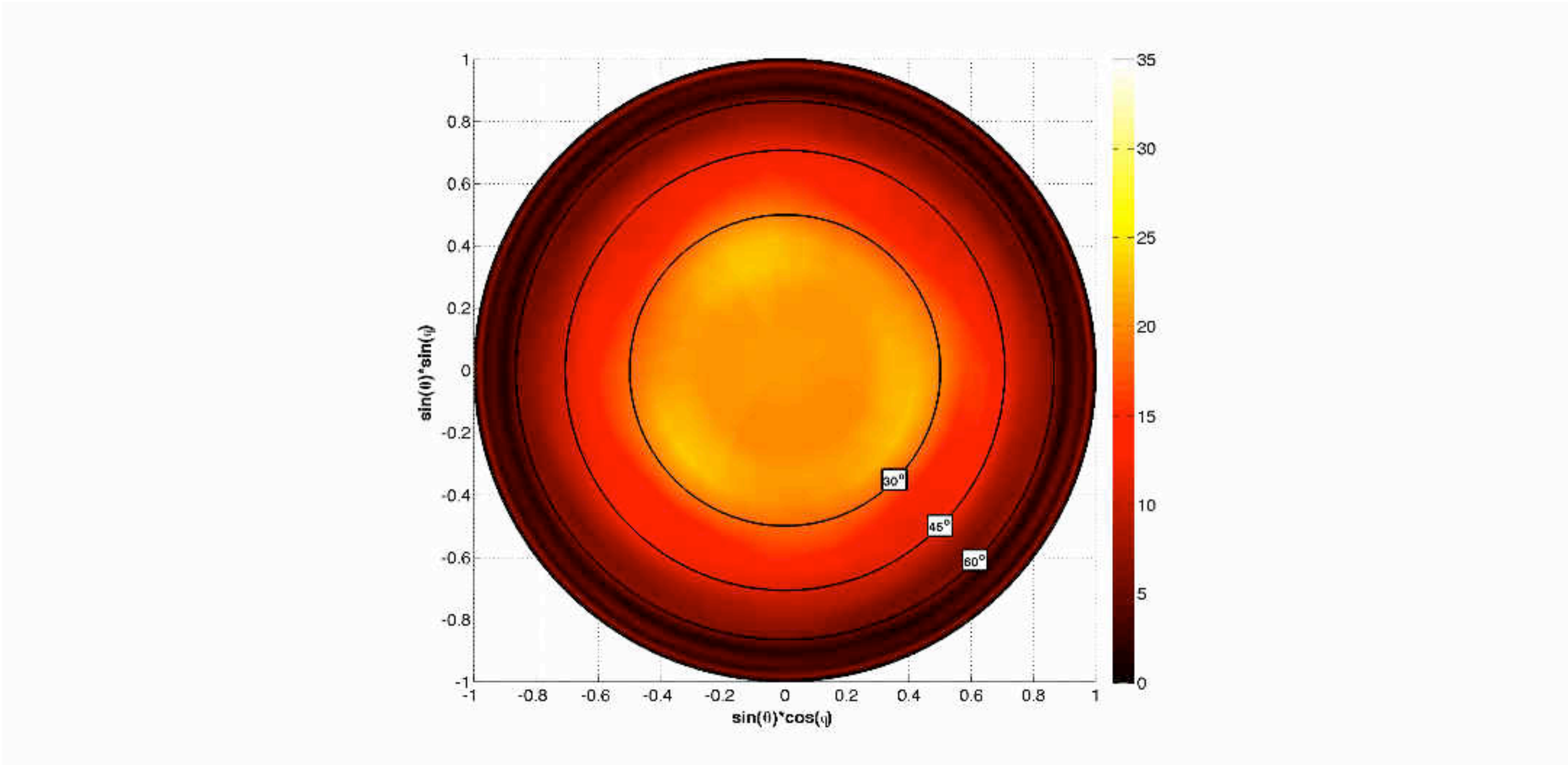

Fig. 13. Intrinsic Cross Polarization Ratio for SKALA in dBs at 50, 70, 100, 200, 300, 400 and 450 MHz (from top to bottom).

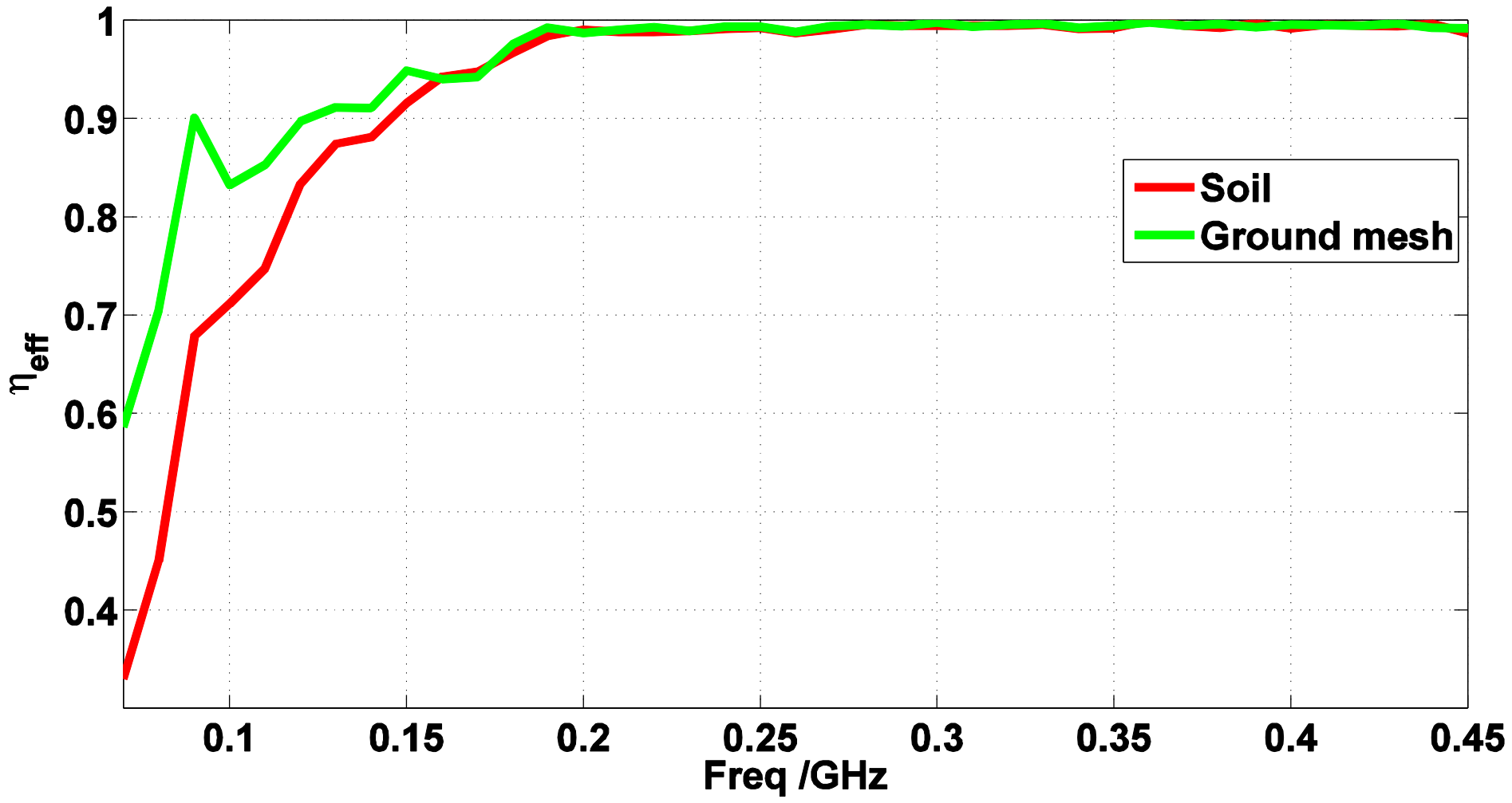

Fig. 14. Radiation efficiency of SKALA.

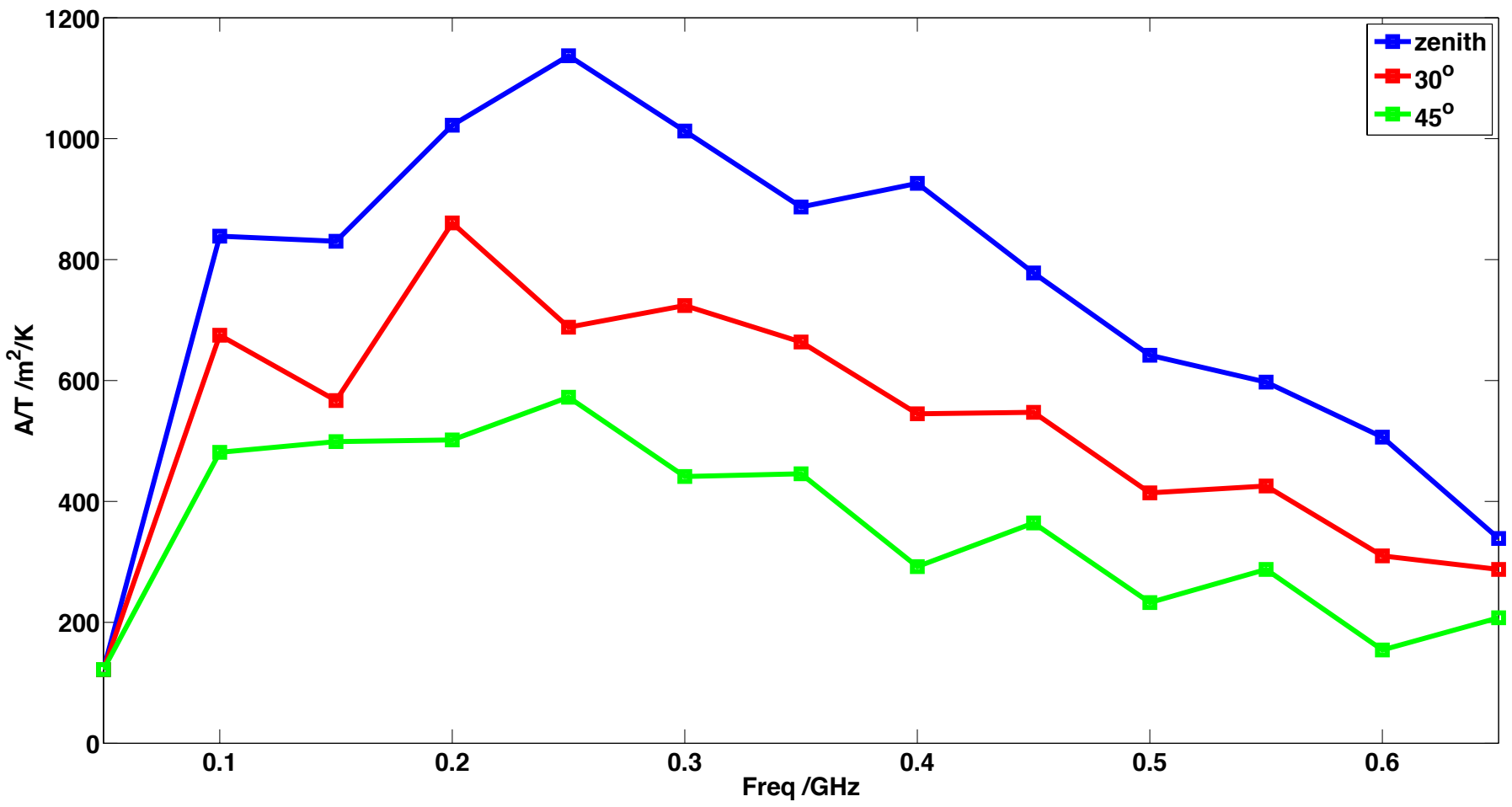

Fig. 15. Sensitivity of $SKA_1$ based on SKALA elements. $T_{sys}$ (in K) is 4,041, 817, 384, 210, 140, 106, 87, 76, 69, 64, 61, 58 and 61 at frequencies from 50 to 650 MHz in 50 MHz intervals.

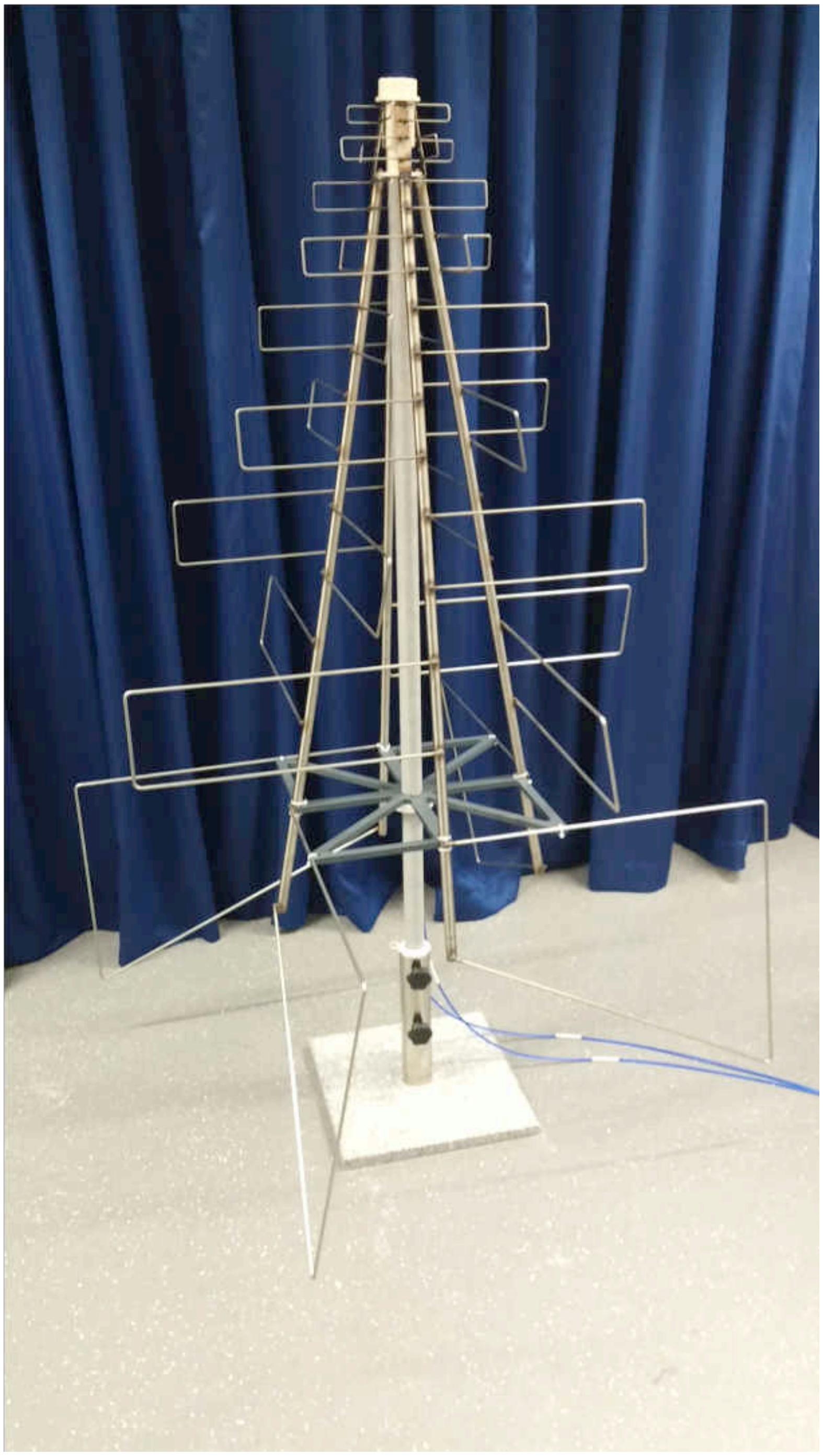

Fig. 16. SKALA.

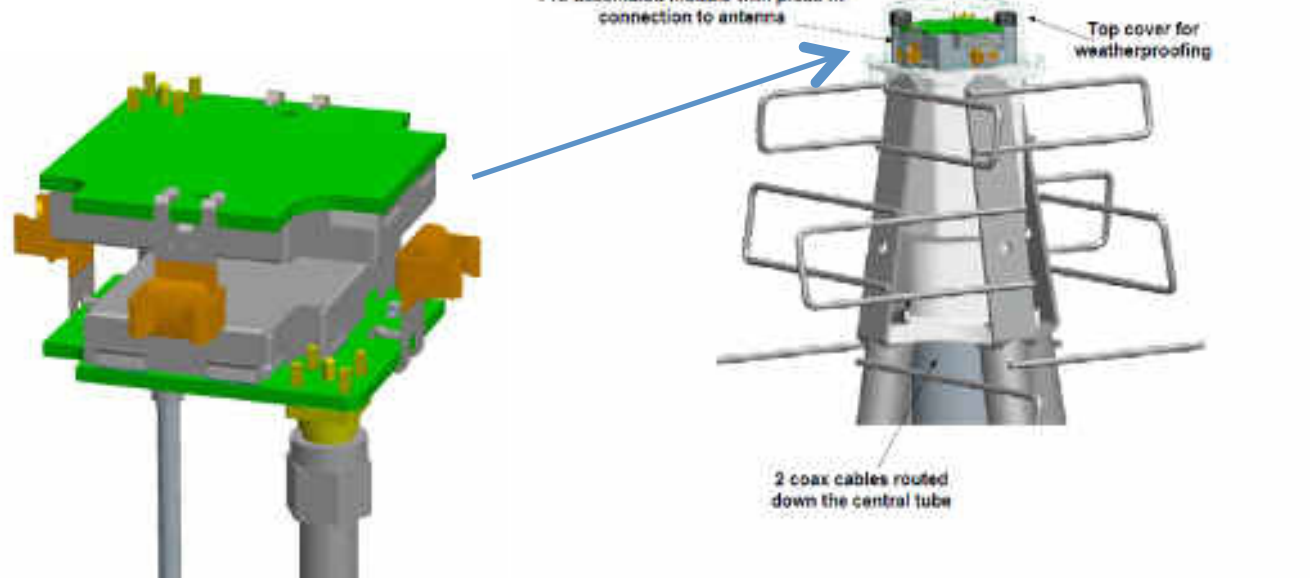

Fig. 17. PCB boards for LNAs located at the top of the antenna.

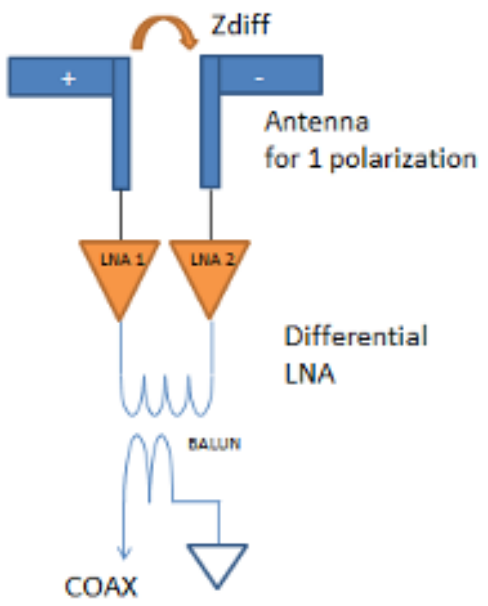

Fig. 18. PCB boards for LNAs located at the top of the antenna.

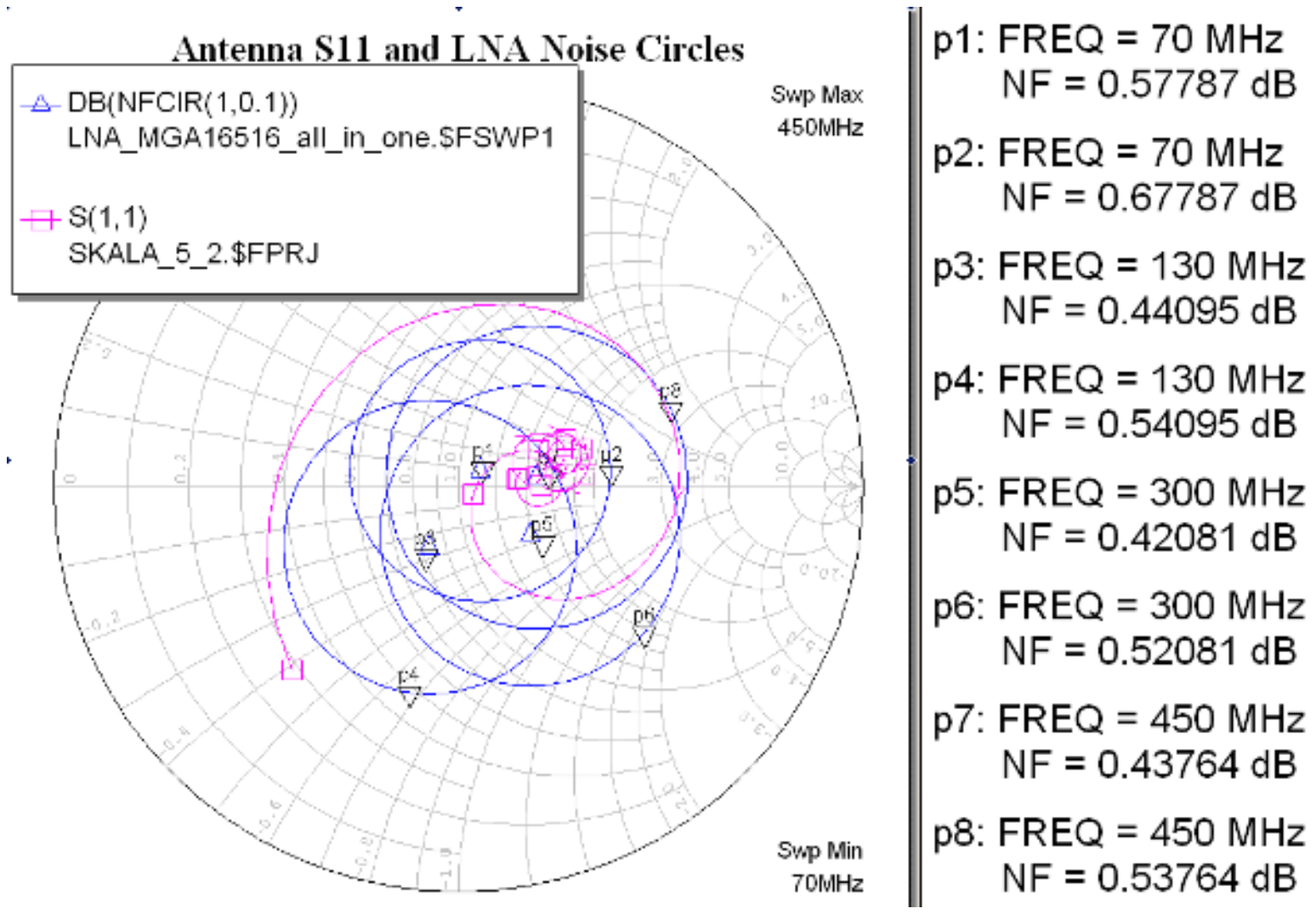

Fig. 19. Antenna impedance and LNA noise circles.

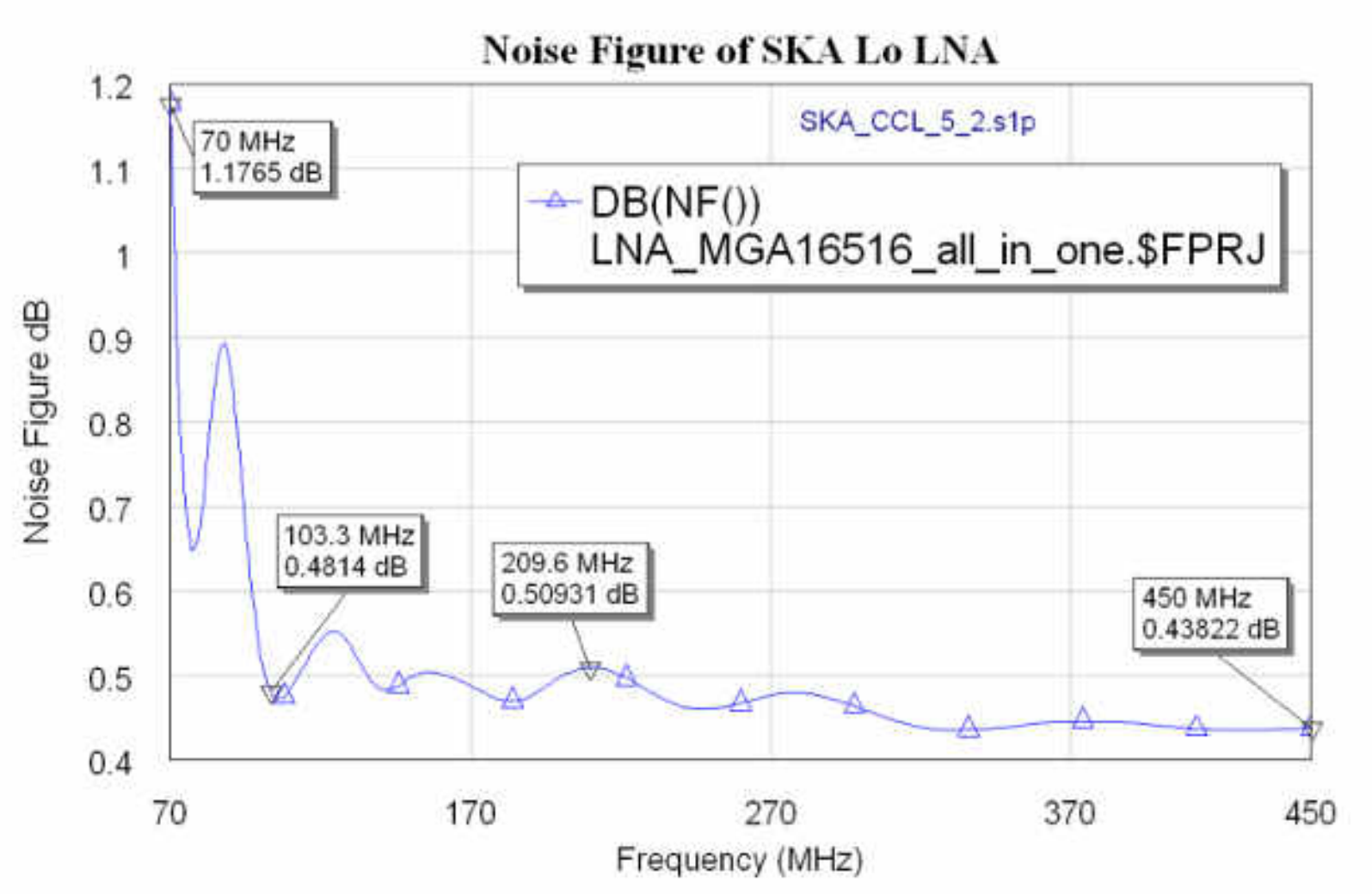

Fig. 20. LNA noise figure when connected to SKALA.

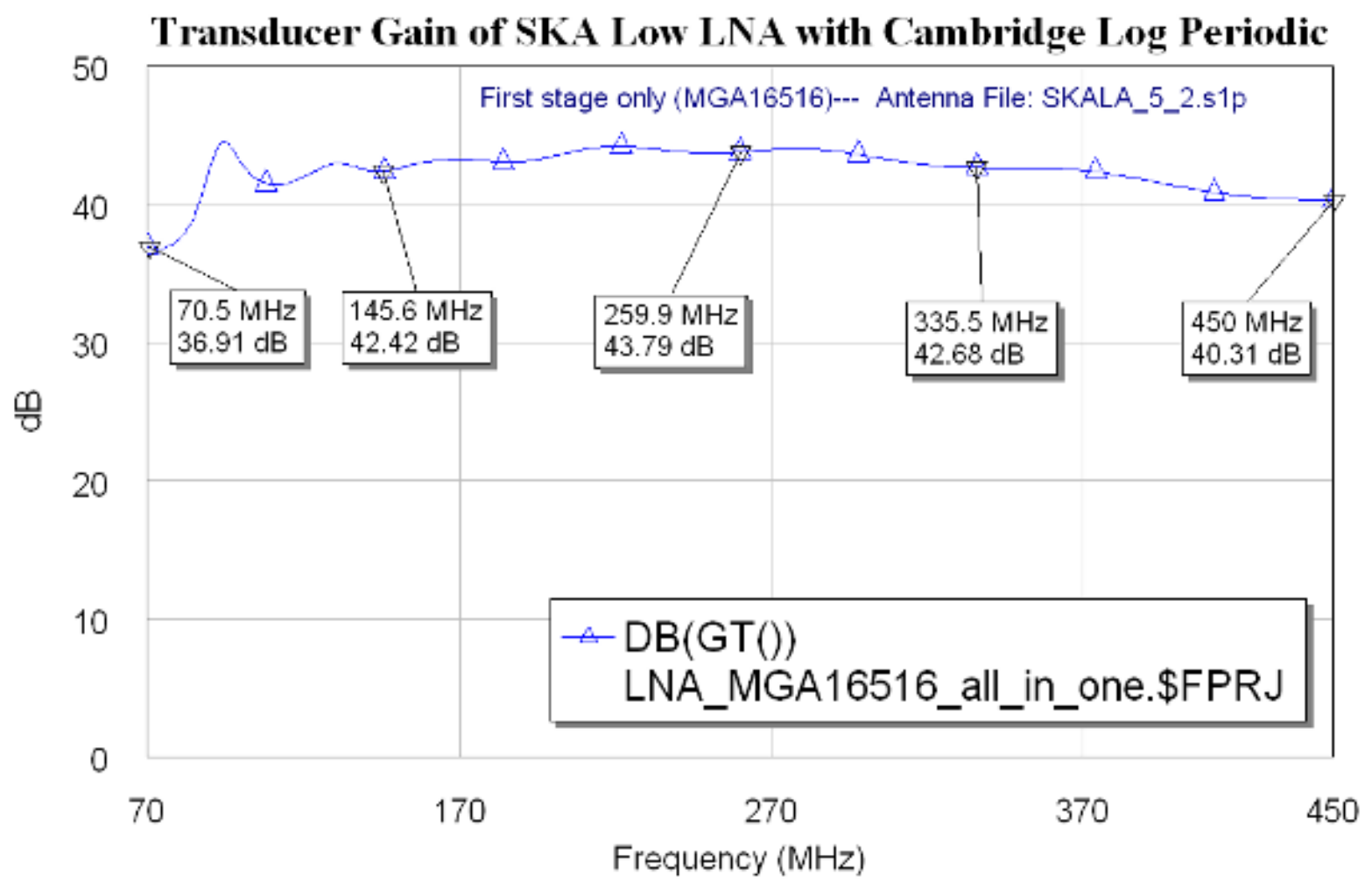

Fig. 21. Transducer gain of the LNA.

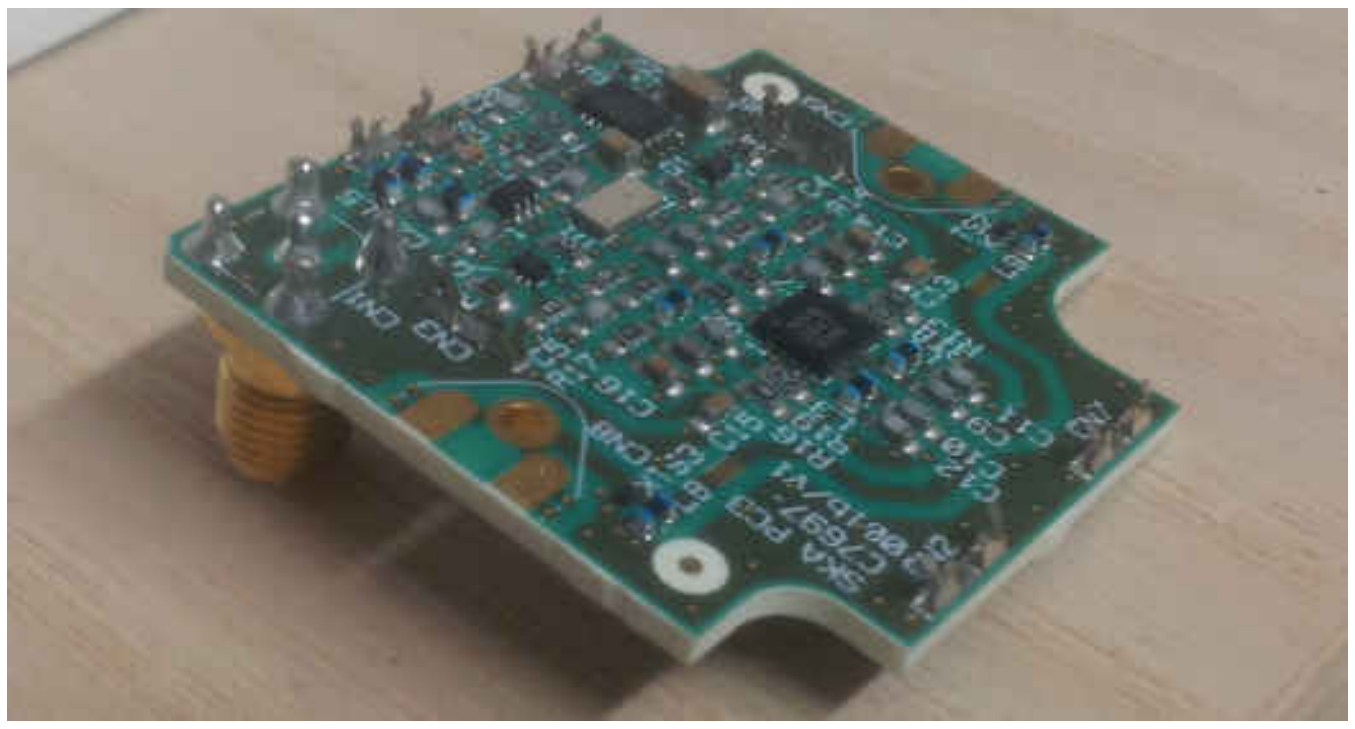
Fig. 22. SKALA's LNA.

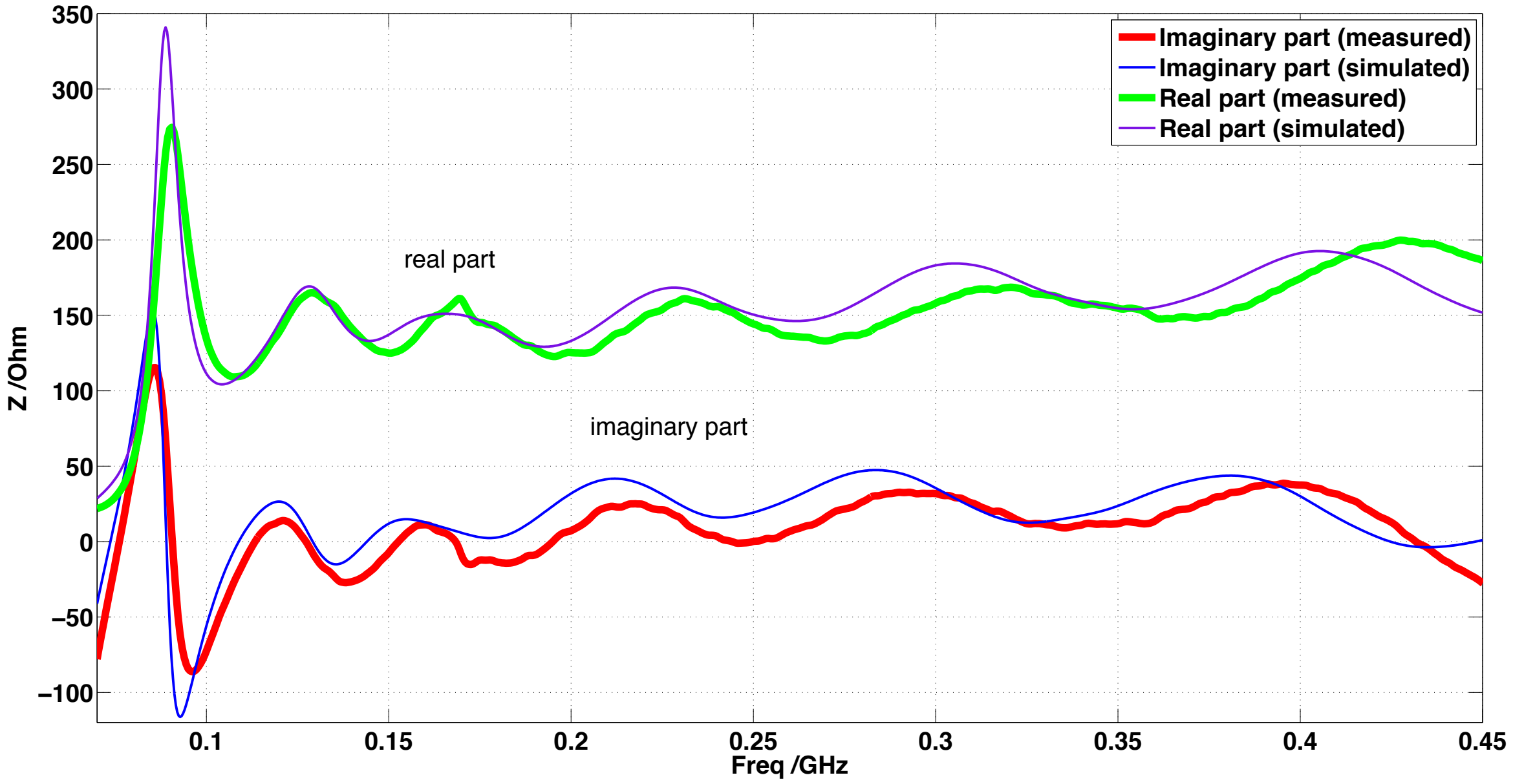

Fig. 23. Impedance of SKALA.

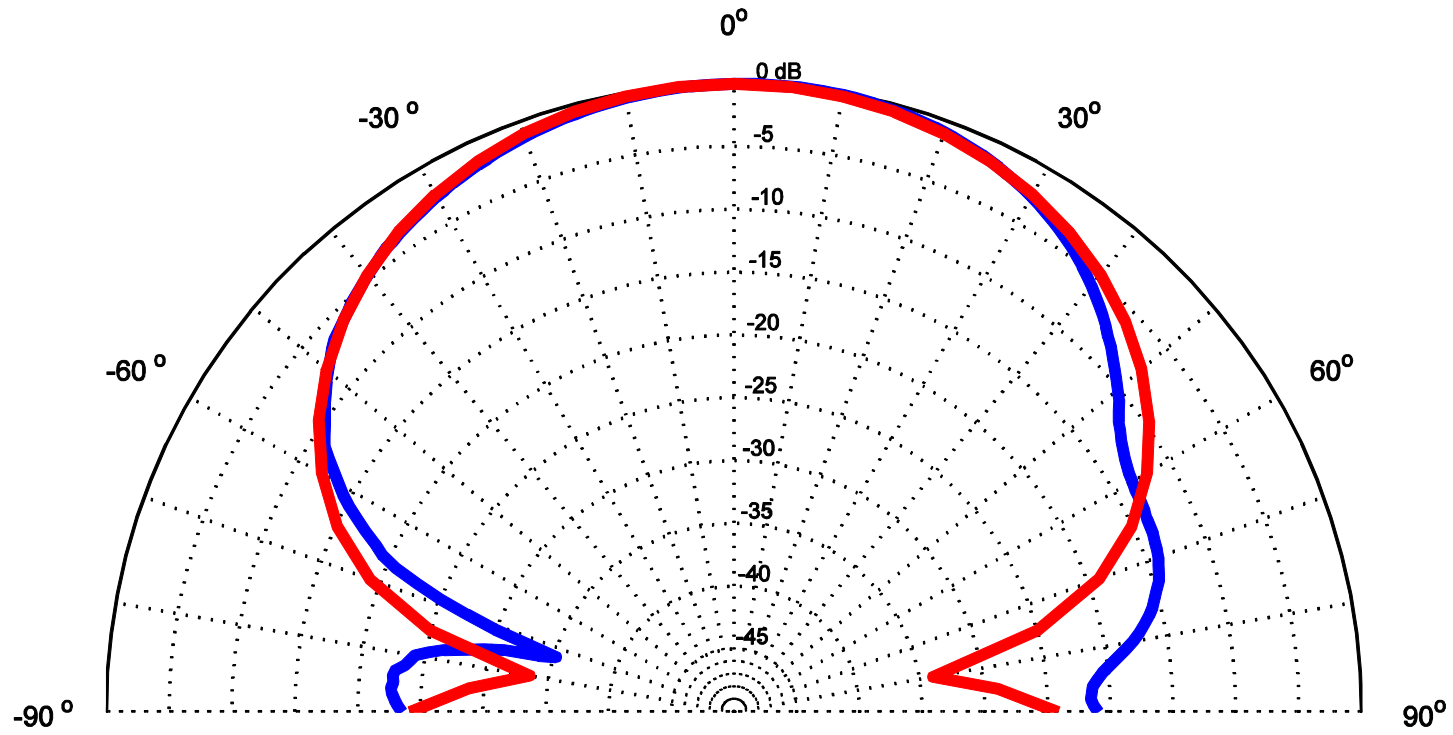

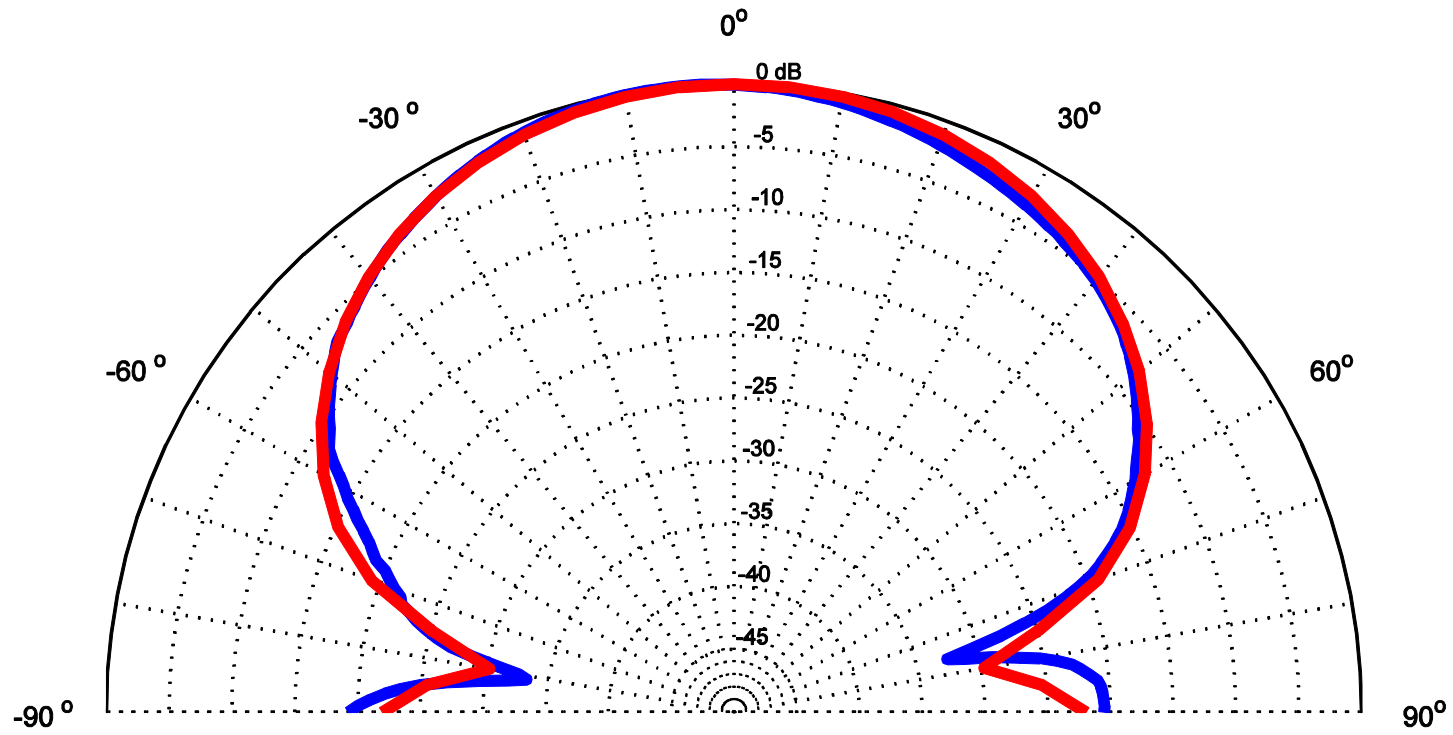

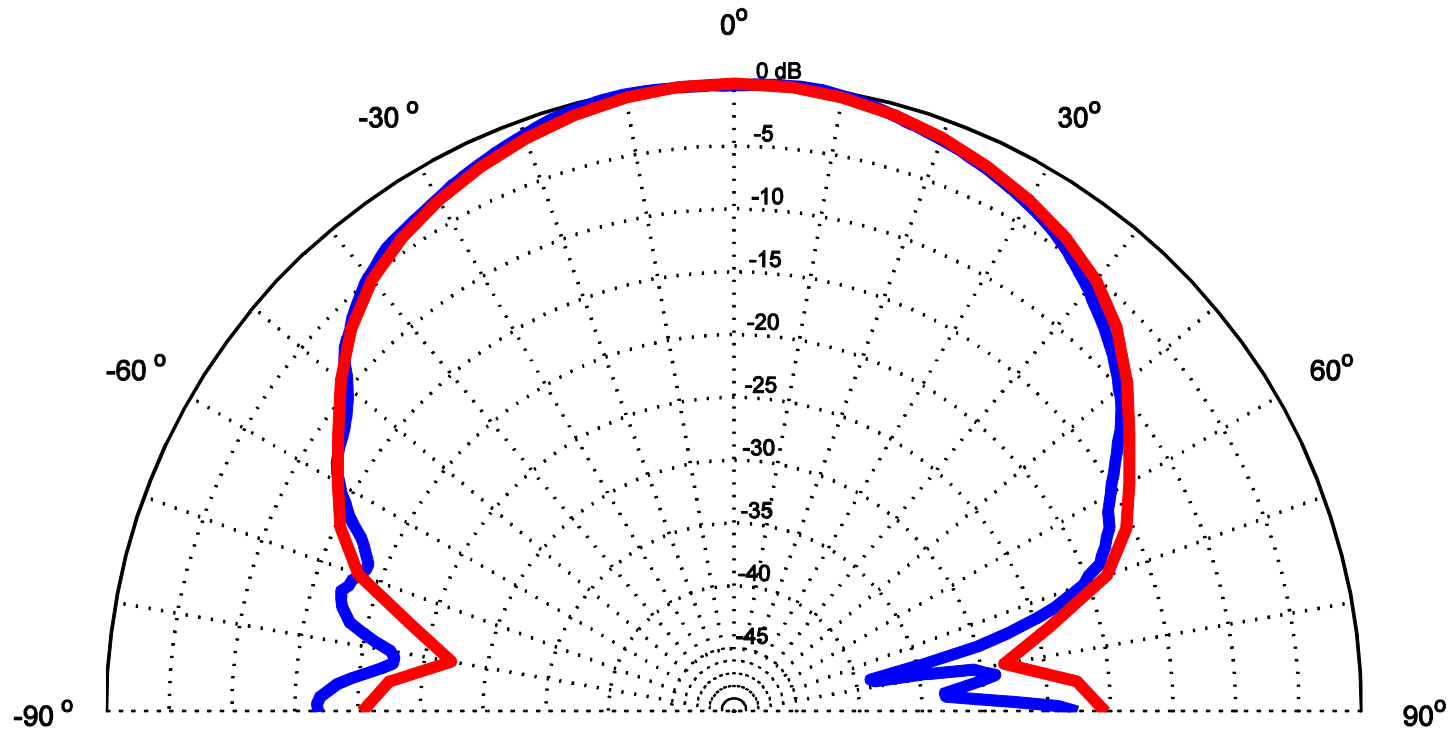

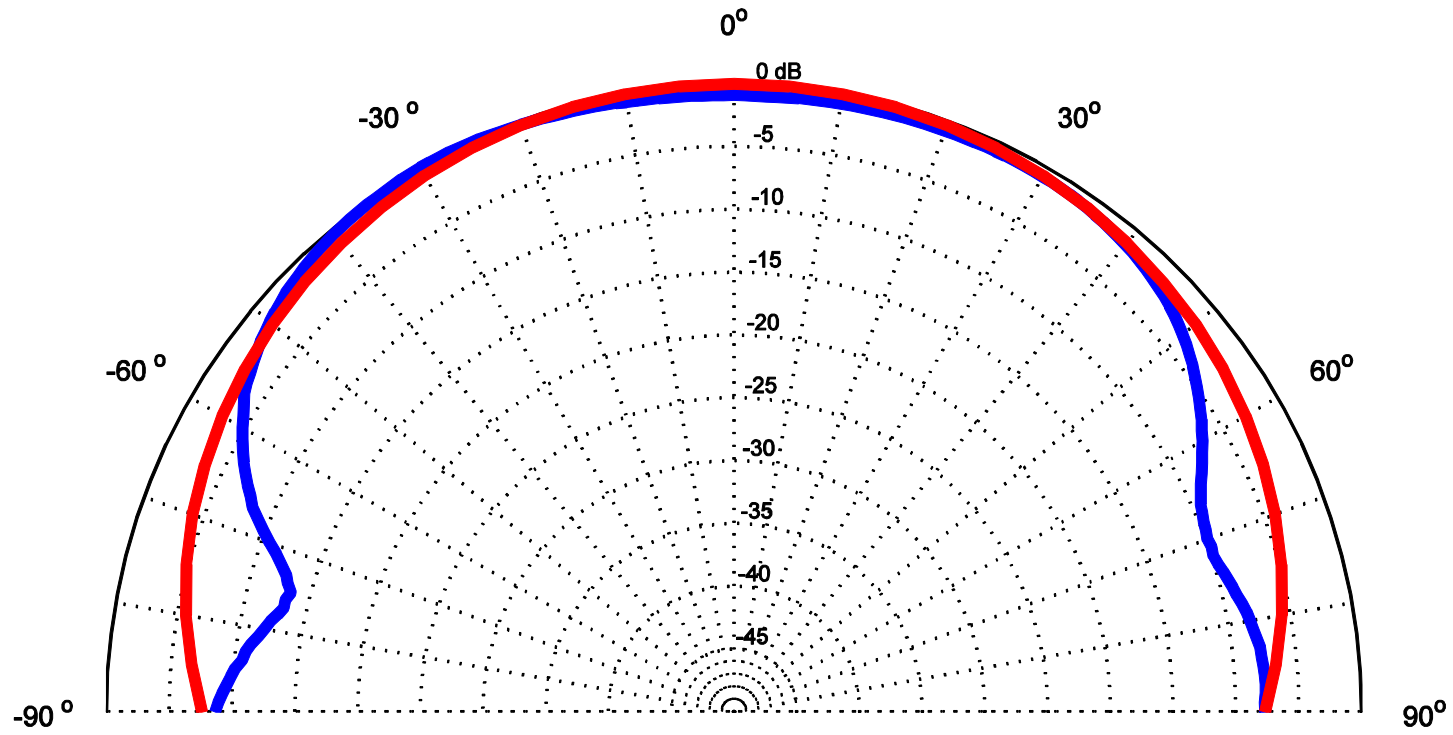

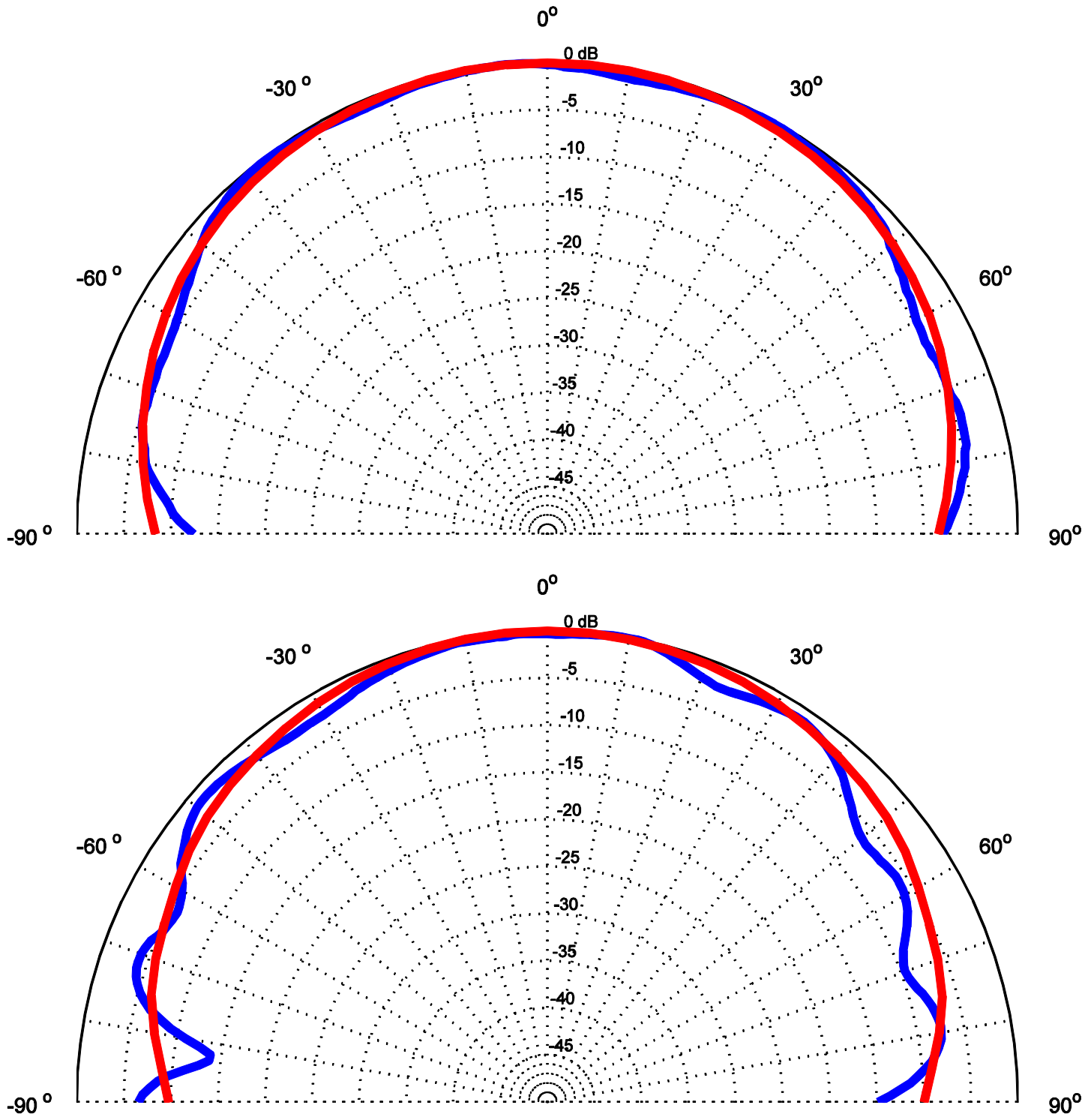

Fig. 24. Measured (blue) and simulated (red) directivity patterns of SKALA on top of a 2x2 m ground plane at 200, 300 and 450 MHz for an E-plane cut (top 3 plots) and an H-plane cut (bottom 3 plots).

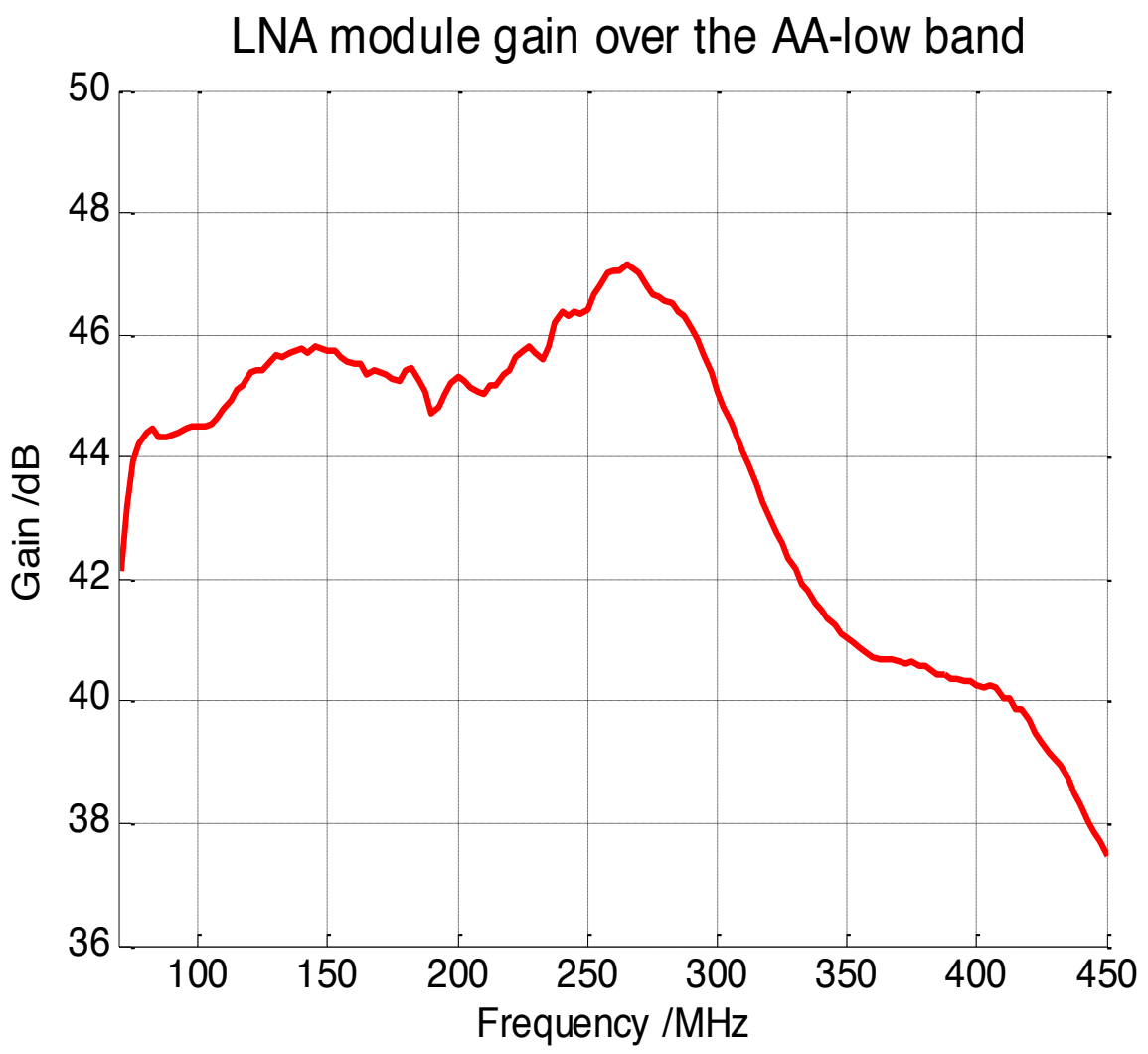

Fig. 25. Gain of LNA.

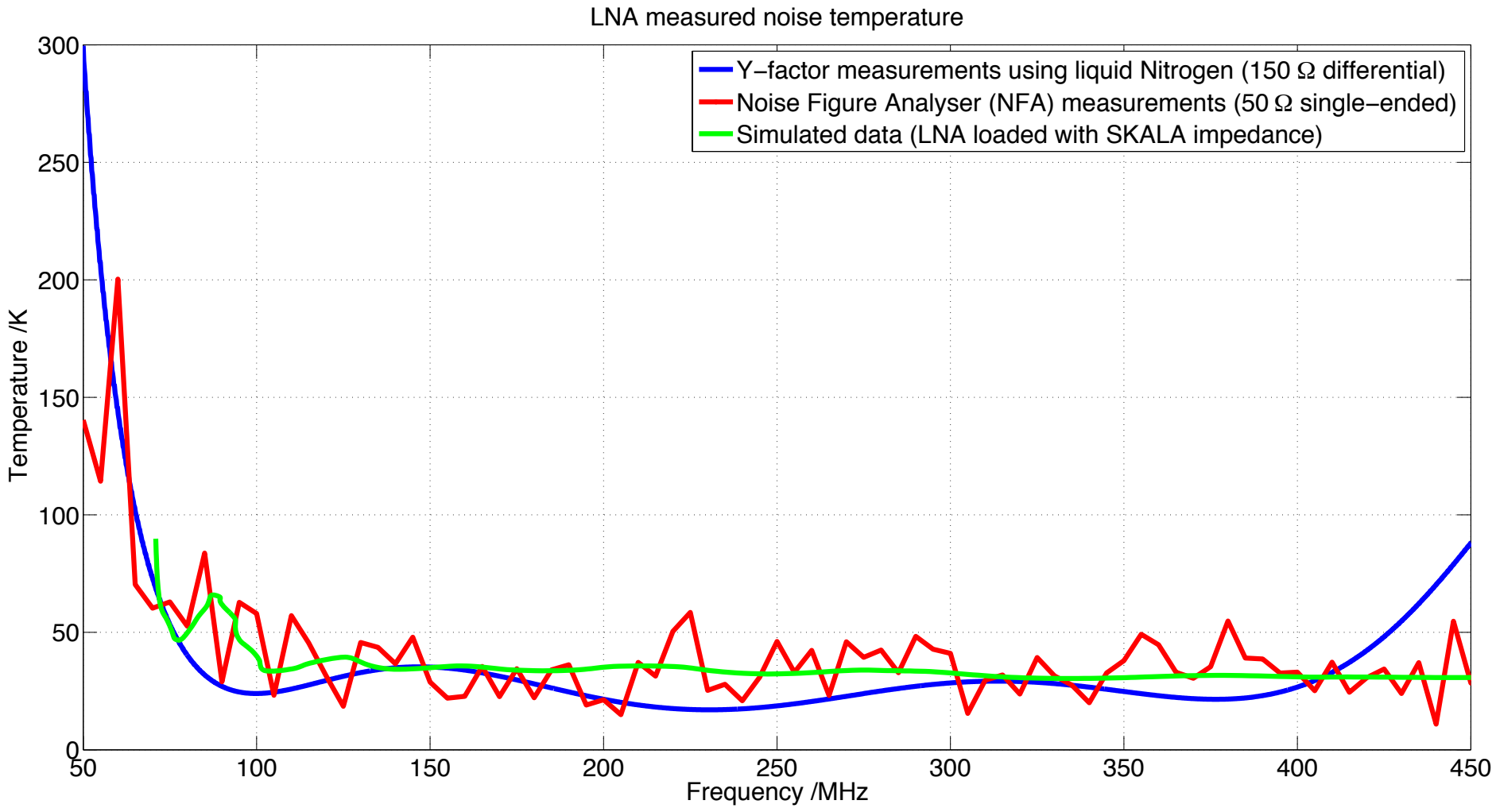

Fig. 26. LNA noise measurement.

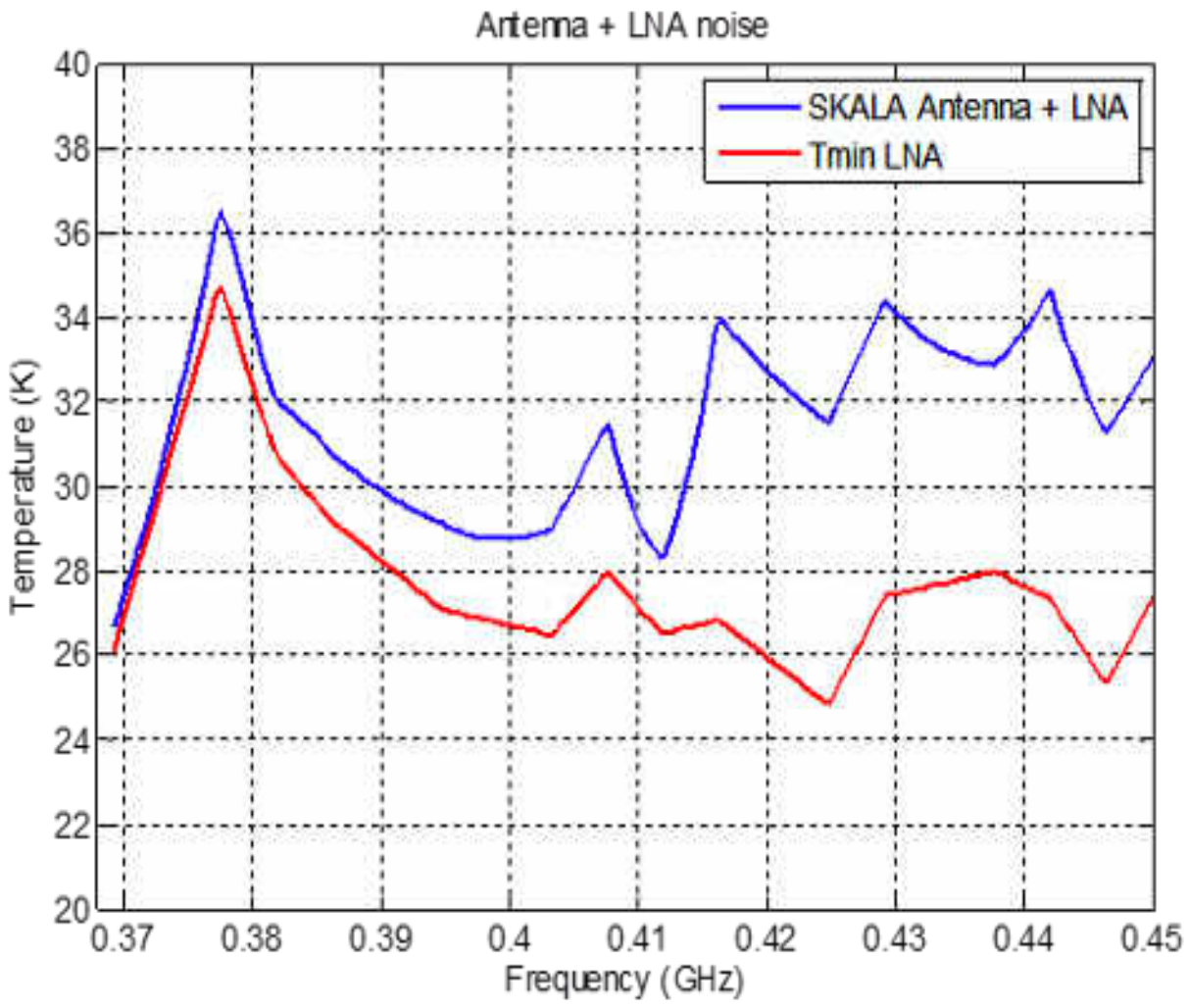

Fig. 27. LNA noise temperature measured with tuner.

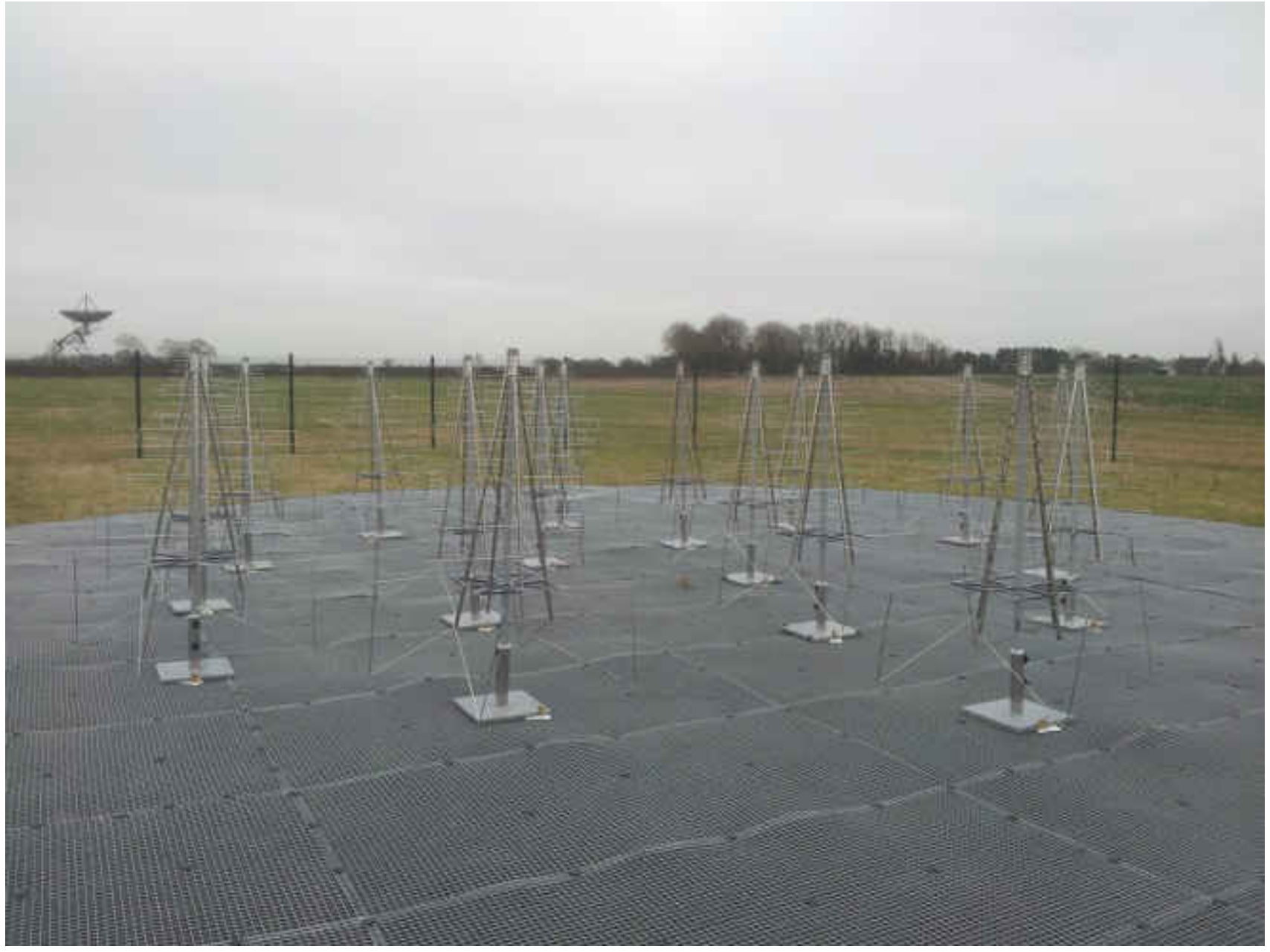

(a)

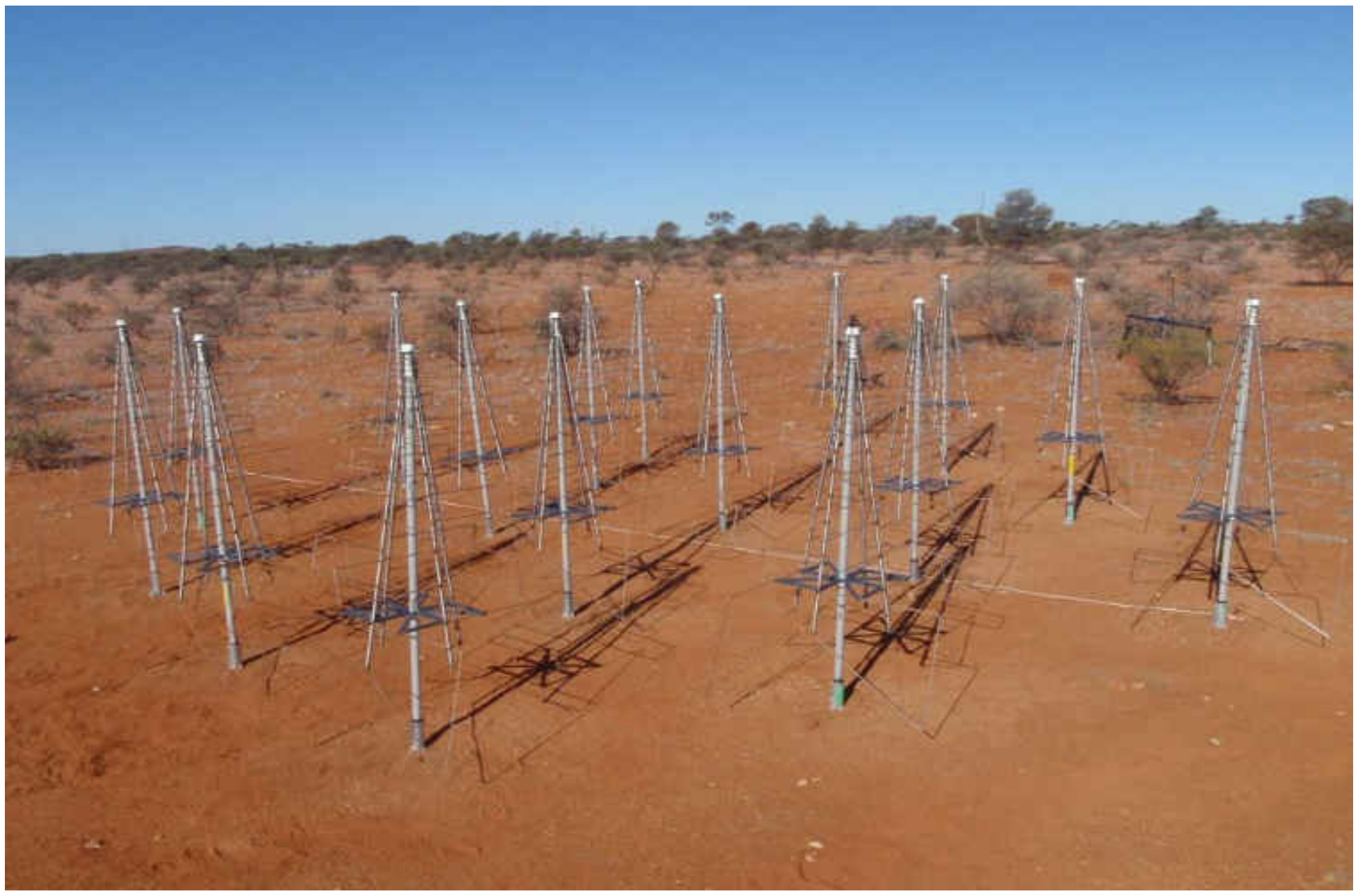

(b)

Fig. 28 (a). AAVS0 array at Cambridge, UK, and (b) AAVS05 array at the MRO site in Australia.

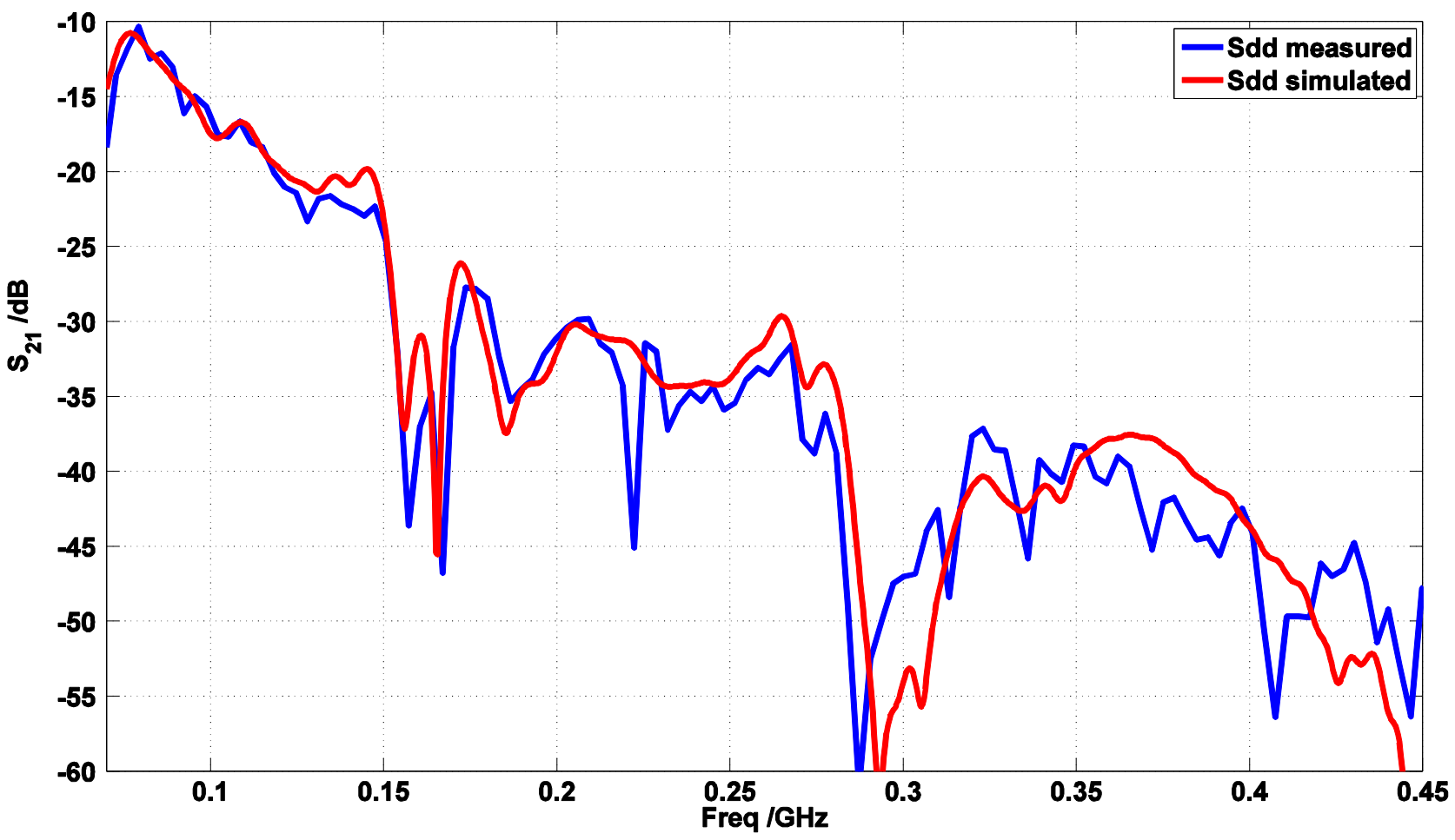

Fig. 29. Coupling between central element and side element in a 3 by 3 regular configuration.

| Dimension | Value |
|---|---|
| *α* | 10º |
| *τ* | 0.76 |
| *G* | 42.5 mm |
| *gnd_D* | 160 mm |
| *H* | 1804 mm |
| *L* | 1371 mm |
| *W* | 34 mm |
| $w_d_2$ | 162 mm |
| $a_d_1$ | 584 mm |
| $b_d_1$ | 680 mm |
| $c_d_1$ | 506 mm |
| *wire thickness (dipoles 1 to 9) as in Fig. 6* | 6, 5, 5, 5, 4, 4, 3, 3 and 3 mm |

Table 1. Main dimensions of SKALA's computer model.